\documentstyle[pre,eqsecnum,aps,multicol,pre]{revtex} 
\input epsf
\input{psfig}
\begin{document} 

\title{Liquid-vapour phase behaviour of a symmetrical binary fluid mixture}

\author{N. B. Wilding}
\address{Department of Physics and Astronomy, The University of Edinburgh,\\
Edinburgh EH9 3JZ, U.K.}

\author{F. Schmid}

\address{Institut f\"{u}r Physik, Johannes Gutenberg Universit\"{a}t, \\
Staudinger Weg 7, D-55099 Mainz, Germany.}

\author{P. Nielaba}

\address{Institut f\"{u}r Physik, Johannes Gutenberg Universit\"{a}t,
\\ Staudinger Weg 7, D-55099 Mainz, Germany.\\ and Institut f\"{u}r
Theoretische Physik, Universit\"{a}t des Saarlands,\\ Postfach 151150,
D-66041 Saarbr\"{u}cken, Germany.}

%\date{May 1995}
\setcounter{page}{0}
\maketitle 
\tighten

\begin{abstract}

Using Monte-Carlo simulation and mean field calculations, we study the
liquid-vapour phase diagram of a square well binary fluid mixture as a
function of a parameter $\delta$ measuring the relative strength of
interactions between particles of dissimilar and similar species. The
results reveal a rich variety of liquid-vapour coexistence behaviour as
$\delta$ is tuned. Specifically, we uncover critical end point
behaviour, a triple point involving a vapour and two liquids of different
density, and tricritical behaviour.  For a certain range of $\delta$, the mean
field calculations also predict a `hidden' (metastable) liquid-vapour
binodal.

\end{abstract}
\thispagestyle{empty}
\begin{center}
PACS numbers 64.60Fr, 64.70.Fx, 05.70.Jk
\end{center}

\begin{multicols}{2}

\section{Introduction} 
\label{sec:intro}

One of the principal deficiencies in our understanding of liquid
mixtures is the nature of the link between the microscopic description
of the system and its macroscopic phase behaviour. For simple single
component fluids, the phase diagram topology is relatively insensitive
to the microscopic properties of the molecules and exhibits (even in
systems with strong anisotropies or long ranged interactions) a
liquid-vapour first order line terminating at a critical point. By
contrast, in binary mixtures, the interplay between the constituent
components leads to a wealth of intriguing phase behaviour
\cite{SCOTT,ROWLINSON} depending on the relative sizes of the molecules
and the strengths of their interactions.

Although the various possible phase diagram topologies have been placed
into a number of categories \cite{SCOTT}, it is not well understood
(even at the mean field level) precisely which microscopic features are
responsible for yielding a given topology. Also unclear is the extent
to which critical fluctuations affect the structure of the phase
diagram {\em i.e.} whether the neglect of correlations in many analytical
theories yields qualitatively (as well as quantitatively) incorrect
phase diagrams.  The task of accurately and reliably deriving the full
phase behaviour of a fluid mixture from knowledge of its microscopic
interactions therefore remains a great challenge.

Evidently, an accurate description of the phase behaviour of a simple
binary fluid model, would provide a useful benchmark against which
current and future liquid-state theories could be tested. In the
present work, we furnish such a description by means of Monte Carlo
simulations of a simple continuum model, the results of which we
compare with mean field calculations.  For reasons of computational
tractability, we consider a {\em symmetrical} binary fluid model, {\em
i.e.} one in which the two pure components $A$ and $B$ are identical
and only the interactions between particles of dissimilar species differ.
Notwithstanding its simplicity, however, the model turns out to reveal
a rich variety of interesting phase behaviour. 

\section{Background}
\label{sec:back}

The phase diagram of a symmetrical binary fluid mixture is spanned by
three thermodynamic fields ($T,\mu,h$), where $T$ is the temperature,
$\mu$ is the overall chemical potential coupling to the total density,
and $h$ is an ordering field coupling to the relative concentrations of
the two fluid components which we assume are allowed to fluctuate.  In
this work we shall restrict our attention to the phase behaviour in the
symmetry plane $h=0$, {\em i.e.} we stipulate that on average the
numbers of $A$ and $B$ particles are equal. Additionally we shall
assume that similar species interactions are energetically more
favourable than dissimilar species interactions. This latter condition
provides for a consolute point (critical demixing transition) at some
finite temperature $T_c$. For temperatures $T<T_c$, there is
coexistence between an $A$-rich liquid and a $B$-rich liquid, while for
$T>T_c$, the system comprises a homogeneous mix of $A$ and $B$
particles. Precisely at $T_c$, the system will be characterised by
strong  critical concentration fluctuations \end{multicols} \newpage
\twocolumn \noindent between the $A$-rich and $B$-rich phases. Such a
demixing transition is analogous to that occurring at the critical
point of a simple spin-$\frac{1}{2}$ Ising model. The difference for an
off-lattice fluid, however, is that the demixing temperature depends on
the density. Consequently one obtains a {\em critical line} of
consolute points $T_c(\rho)$ [or $T_c(\mu)]$, which is commonly
referred to as the `$\lambda$ line'.

In addition to exhibiting consolute critical behaviour, binary fluids
can also exhibit liquid-vapour (LV) coexistence, in much the same way
as does a single component fluid. It transpires, however, that the LV phase behaviour of binary
mixtures is considerable richer than that of simple fluids. This
difference is traceable to the additional ingredient of concentration
fluctuations, which couple to the density fluctuations and which can
radically alter the LV phase behaviour.  Since concentration
fluctuations are strongest on the $\lambda$ line, one expects that
alterations to the LV coexistence behaviour will be greatest where this
line approaches the LV coexistence curve. 

Perhaps not surprisingly, binary fluids mixtures are not the only fluid
systems in which first order phase coexistence behaviour is influenced
by the proximity of a critical line. The earliest sightings of such
effects appears to have been in analytical studies of various
lattice-based fluid models \cite{HALL,HEMMER1,STELL}. Some time later,
a detailed Landau theory study of a model for sponge phases in
surfactant solution \cite{ROUX}, revealed a rich variety of first order
phase behaviour as the path of the $\lambda$ line was varied. More
recently, similar behaviour was uncovered in extensive mean field and
density functional theory investigations of a number of symmetrical
continuum fluid models, namely the classical Heisenberg spin fluid
\cite{HEMMER,OUKOUISS,TAVARES,WEIS}, a dipolar fluid model
\cite{ZHANG,GROH}, and the van der Waals-Potts fluid \cite{ZALUSKA}.

\begin{figure}[h]
%\vspace*{-2mm}
\setlength{\epsfxsize}{6.5cm}
\centerline{\mbox{\epsffile{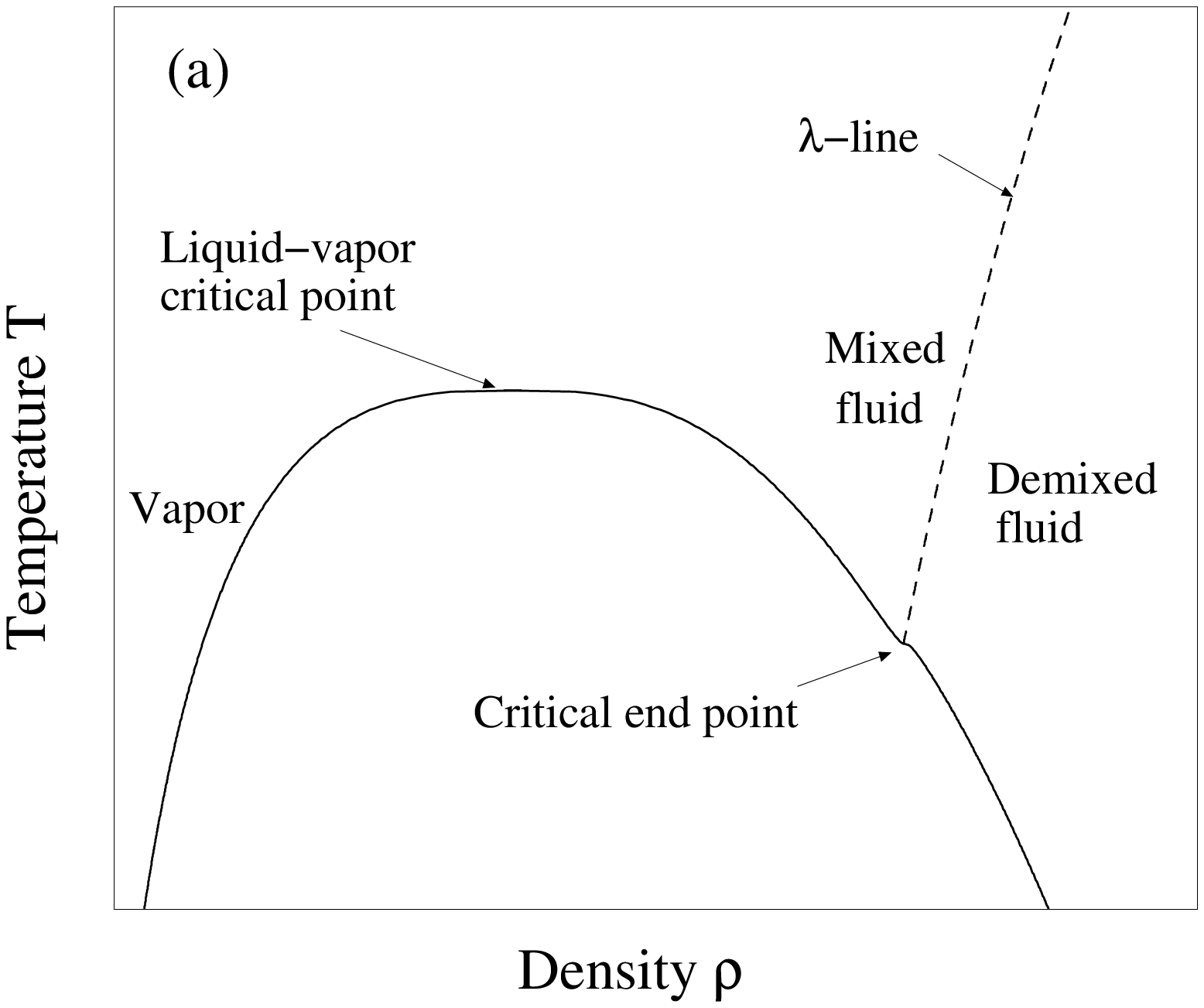}}} 
\vspace*{5mm}
\centerline{\mbox{\epsffile{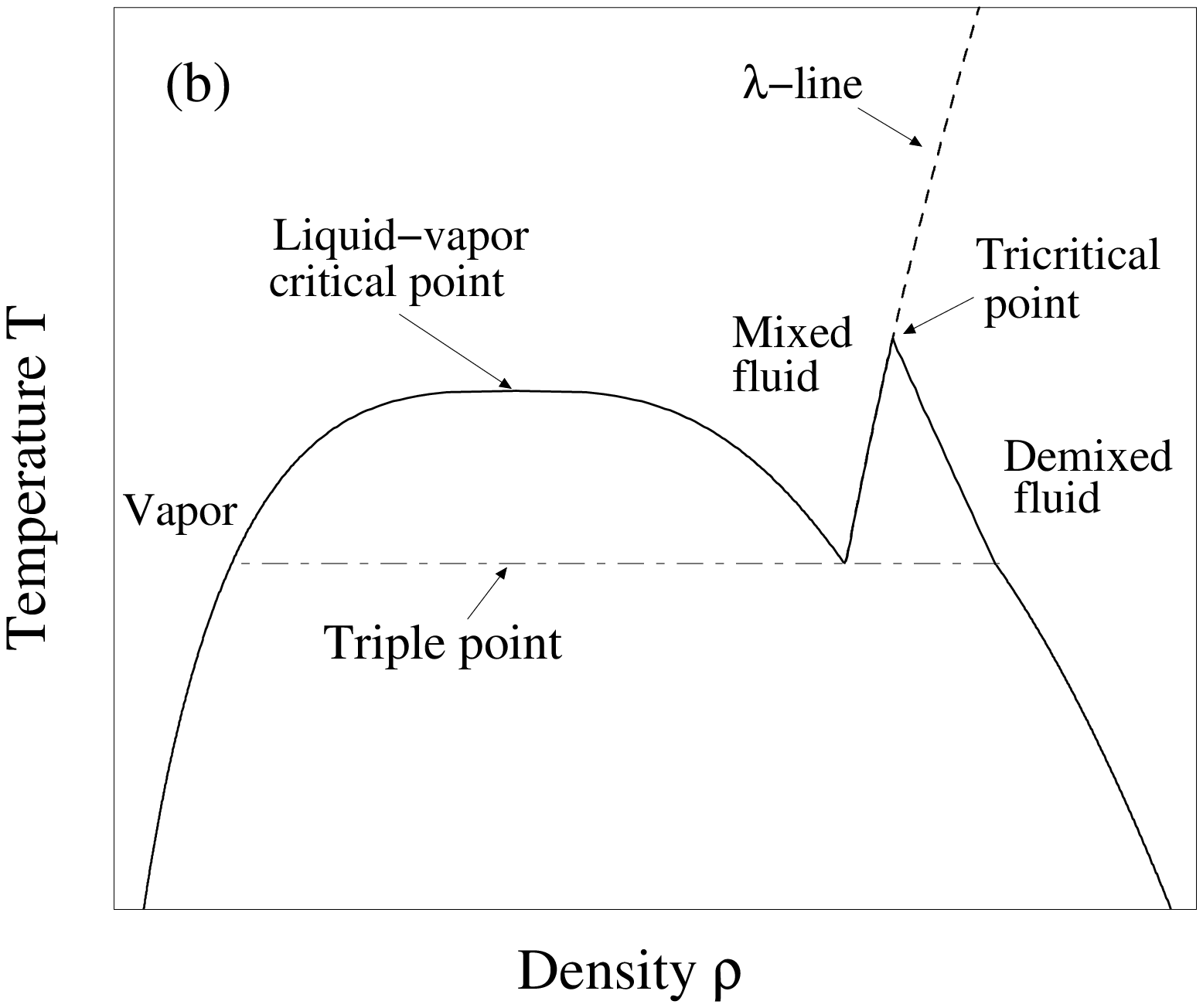}}} 
\vspace*{5mm}
\centerline{\mbox{\epsffile{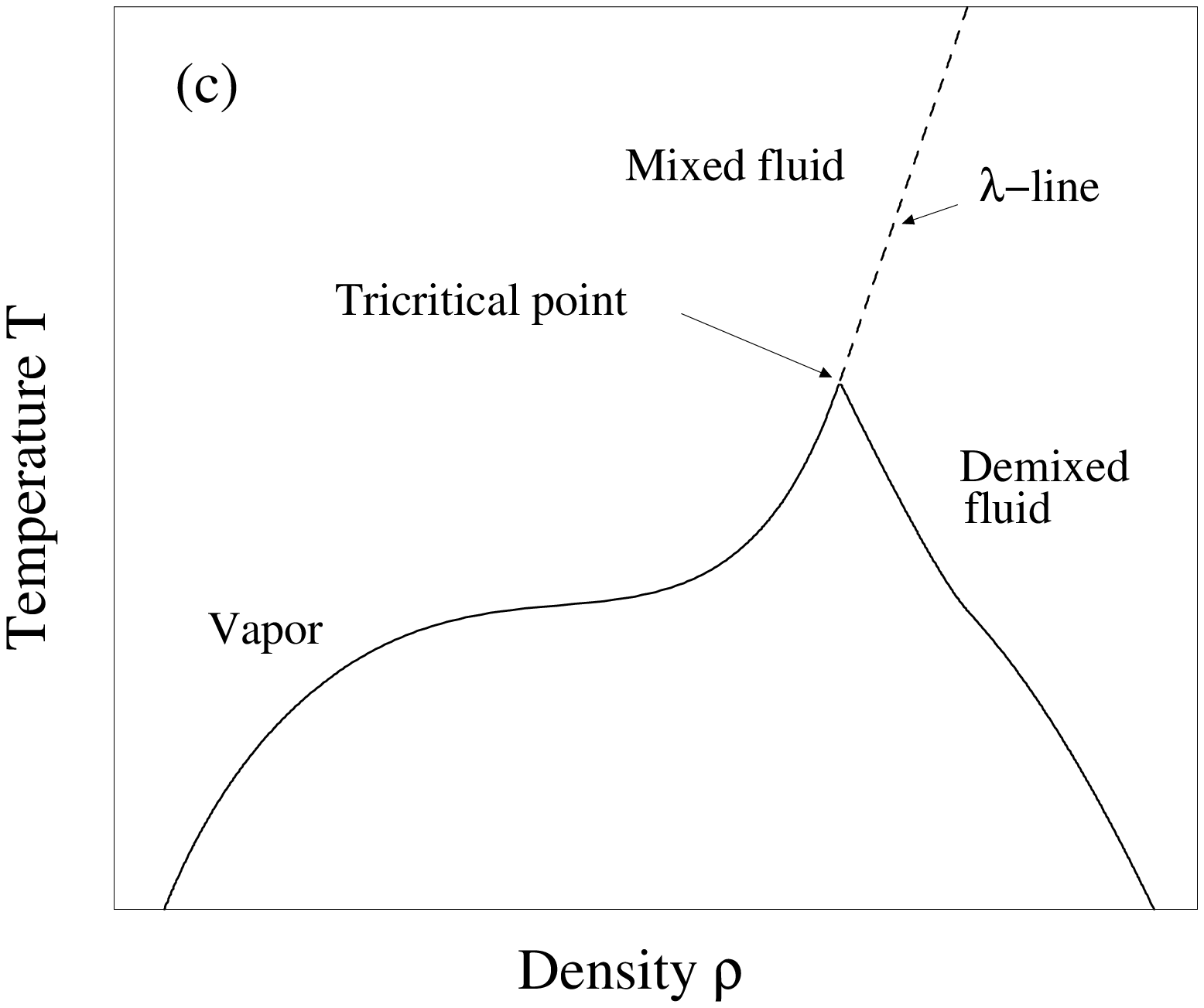}}} 
\vspace*{5mm}
\caption{Schematic representation of the three types of phase diagram
for a symmetrical binary fluid mixture in the density-temperature
plane, as described in the text. The full curve is the first order 
liquid-vapour coexistence envelope, while the dashed curve is the
$\lambda$ line of critical demixing transitions.}

\label{fig:schem}
\end{figure}

Despite dealing with ostensibly quite distinct models, the gross
features of the mean field phase behaviour emerging from these studies
appears to be essentially model-independent. This behaviour is
illustrated schematically in fig.~\ref{fig:schem} and involves three
possible LV phase diagram topologies, depending on the path of the
critical line relative to the LV line.  To describe this behaviour we
shall employ the language of the symmetrical binary fluid. In so doing,
we anticipate the result of sections~\ref{sec:mf} and~\ref{sec:mc},
namely that the same scenario is played out in this case too. Of
course, to obtain the corresponding behaviour for other systems, eg.
the magnetic or dipolar fluids, one need only substitute the
appropriate nomenclature eg. `mixed fluid' $\to$ `paramagnetic fluid'.

%In the symmetrical binary fluid, it transpires that the locus of the
%$\lambda$ line relative to the LV line may be varied by tuning the
%parameter $\delta=J_/J_l$ representing the relative strength of
%interactions between similar and dissimilar particles. 

Fig.~\ref{fig:schem}a depicts schematically the mean field phase
diagram obtained when the model parameters are chosen such that the
$\lambda$ line approaches the first order phase boundary well below the
liquid-vapour critical point. In such a situation, the $\lambda$ line
{\em intersects} the LV line at a critical end point (CEP). At the CEP,
a critical liquid coexists with a non-critical vapour. Below the CEP
temperature one finds a triple line in which a vapour coexists with an
$A$-rich liquid and a $B$-rich liquid. Owing to the symmetry, these two
liquids have the same density.

Alternatively, for different model parameters, the $\lambda$ line may
intersect the LV line at the liquid-vapour critical point
[fig~\ref{fig:schem}c]. Under such conditions,  phase coexistence
between the vapour and the mixed fluid is preempted by the demixed fluid
phase. One then obtains a tricritical point \cite{NOTE1,LAWRIE} in which
three phases (a vapour, an $A$-rich liquid, a $B$-rich liquid)
simultaneously become critical. 

The intermediate situation is shown in fig.~\ref{fig:schem}b, and
occurs when the $\lambda$ line approaches the LV line at a temperature
somewhat (but not greatly) below the liquid-vapour critical
temperature. In this case, the phase diagram combines the features of
the previous two cases. One finds a {\em triple point} in which a
vapour coexists with a mixed liquid at intermediate density and a
demixed liquid of higher density \cite{NOTE3}. Above the triple point
temperature, a demixed vapour and a demixed liquid coexist at low and
moderate densities, becoming identical above the liquid-vapour critical
point. At higher densities, a mixed liquid and the demixed liquid
coexist, becoming identical at a tricritical point.

That the scenario described above is generic to a range of apparently
distinct fluid models (eg. dipolar, magnetic and binary fluids), is
perhaps slightly surprising at first sight. On closer examination,
however, it becomes clear that the model differences are only skin-deep.
All the systems in which this behaviour has yet been identified, can, in
essence,  be regarded as fluids in which each particle carries an
internal degree of freedom,  eg. a spin or dipolar moment. The
symmetrical binary fluid model shares this behaviour because the
particle species label is analogous to a two-state `spin' variable.

Notwithstanding the substantial body of analytical evidence supporting
the scenario of phase behaviour shown in fig.~\ref{fig:schem}, it must
necessarily be regarded as somewhat tentative given the notorious
inability of mean field theories to account accurately for critical
behaviour below the upper critical dimension.  In view of this,
independent corroboration by computer simulation is clearly desirable
and necessary. In recent times, such studies have indeed started to
appear. 

The first simulations to study the confluence of a critical demixing
line and a first order LV line sought to elucidate the behaviour when
the intersection occurs at the LV critical point, {\em i.e.} at a tricritical
point [cf. fig.~\ref{fig:schem}c]. Investigations of a two-dimensional
spin fluid \cite{WILDING2} demonstrated that the  tricritical
point properties are identical to those of the 2D Blume-Capel model, as
one might expect on universality grounds. Other studies, this time
focusing on the 3D classical Heisenberg fluid \cite{WEIS,LOMBA} and
spin-$\frac{1}{2}$ quantum fluids \cite{MARX,SENGUPTA,DESMEDT}, mapped the LV phase
envelope and the $\lambda$ line around the tricritical point and
compared the results with mean field calculations. Modest agreement was
found.

As regards the situation shown in fig.~\ref{fig:schem}a and
~\ref{fig:schem}b, there is a still a paucity of simulation
data. This is presumable traceable to the practical difficulties
associated with studying first order phase coexistence deep within the
two phase region. The basic problem is the ergodic (free energy)
barrier to sampling both coexisting phases in a single simulation. This
barrier arises because the phase space path leading from one pure phase
to another necessarily passes through interfacial configurations of
large free energy, having a concomitantly low statistical weight. Such
configurations occur only very rarely in a standard Monte Carlo (MC)
simulation. Fortunately, however, a recently introduced biased-sampling
technique known as multicanonical preweighting \cite{BERG} allows one
to negotiate this barrier, and thus obtain accurate estimates of
coexistence properties \cite{GEMC}. The efficacy of the method for
investigating fluid phase coexistence was first demonstrated in
\cite{WILDING3}. Very recently, it has also been employed to study 
critical end point behaviour in a Lennard-Jones binary fluid model
\cite{WILDING1}, [cf. fig.~\ref{fig:schem}a].  In this study, the
intersection with the $\lambda$ line was shown to engender a
singularity in the first order phase boundary--in accord with earlier
theoretical predictions \cite{FISHER}. For the LV phase envelope,  this
singularity is manifest as a bulge in the liquid branch density (as
indicated schematically in fig.~\ref{fig:schem}a).

Thus while fragments of the picture of LV behaviour in symmetrical
fluids have been set in place by simulation, much clearly remains to be
done before a reliable and comprehensive overview emerges. In
particular, no evidence has yet been reported for the existence of the
triple point behaviour shown schematically in
fig.~\ref{fig:schem}b.  There has also been no systematic
simulation study of the full range of LV phase behaviour for a {\em
single} model. Consequently little reliable information exists
concerning the manner in which one type of phase diagram evolves into
another. 

In the present work we have attempted to remedy this situation by
performing a systematic MC simulation study of the LV phase behaviour
of a binary fluid mixture. The model studied has an interparticle
potential of the square-well form:

\begin{eqnarray}
U(r)= &\infty \hspace{1cm} & r< \sigma \nonumber \\
U(r)= & -J \hspace{1cm} & \sigma \leq r< 1.5\sigma \nonumber \\
Ur)    = & 0 \hspace{1cm}   & r\geq 1.5\sigma 
\label{eq:potdef}
\end{eqnarray}
Here $r$ is the particle separation, $J$ is the well depth or
interaction strength, and $\sigma$ is the hard-core radius. In general
there will be a number of different interaction strengths $J^{AA},
J^{BB}, J^{AB}$ depending on the respective species of the interacting
particles. However, in the symmetrical case with which we shall be
concerned, one has simply

\begin{eqnarray}
J^{AA}(r)  = & J^{BB}(r)   & =J(r)    \nonumber   \\
J^{AB}(r)  = & \delta J(r) . & 
\label{eq:symmdef}
\end{eqnarray}
The parameter $\delta=J_{AB}/J\leq 1$  determines the degree to
which interactions between dissimilar species are less favourable than
similar species interactions. Since $\delta$ is the only free model
parameter, it controls the complete range of possible phase behaviour.

Using MC simulation we obtain the liquid-vapour phase behaviour of this
system for a range of values of $\delta$. The results are compared with
explicit mean field model calculations in order to assess the latter's
ability to reproduce the actual phase behaviour. Landau theory
calculations are also reported which furnish physical insight into the
general mechanisms by which the coupling of density and concentration
fluctuations engender the various types of observed phase behaviour.
Additionally we calculate (both explicitely for our model and within
Landau theory) the spinodal lines of the phase diagram. This reveals
that for a certain range of $\delta$ there exists a `hidden' or
metastable binodal, the presence of which is expected to have
implications for the system dynamics. 

The remainder of our paper is organised as follows.  In
section~\ref{sec:mf} we detail the explicit mean-field and general
Landau theory calculations for the phase diagram as a function of
$\delta$. The Monte Carlo simulation results for the liquid-vapour
phase behaviour are presented in section~\ref{sec:mc}. Finally, in
section~\ref{sec:concs} we conclude by comparing the simulation and
mean field results, and discussing prospects for future work.

\section{Mean-field calculations}
\label{sec:mf}

In this section we present two complementary mean field studies of
symmetrical binary mixtures. The first is an explicit investigation of
the square-well binary mixture model, also studied by simulation in
section~\ref{sec:mc}. The second is a general Landau theory treatment
aimed at obtaining physical insight into the origin of the various
phase diagram topologies.

\subsection{Explicit model calculations}
\label{sec:mfexp}

Let us consider a binary fluid in which the two particle species $A$
and $B$ interact via the symmetrical square-well potential of
equations~\ref{eq:potdef} and~\ref{eq:symmdef}. For analysis purposes
it will prove useful to decompose this potential into a hard-sphere
component plus a short ranged attractive part. To this end we rewrite
equation~\ref{eq:potdef} as

\begin{mathletters}
\begin{equation}
U(r)=U_{HS}(r)+J(r) ,
\label{eq:pot1}
\end{equation}
where 
\begin{eqnarray}
U_{HS}(r)= &\infty \hspace{1cm} & r< \sigma \nonumber \\
U_{HS}(r)    = & 0 \hspace{1cm}     & {\rm otherwise}
\label{eq:pot2}
\end{eqnarray}
and
\begin{eqnarray}
J(r) =& -J \hspace{1cm} & \sigma \le r \le 1.5\sigma \nonumber   \\
J(r) =& 0  \hspace{1cm} & {\rm otherwise}.
\label{eq:pot3}
\end{eqnarray}
\end{mathletters}

For a binary fluid, the interaction between two particles depends
on their respective species.  To deal with this we introduce a
two-state species variable $s_i$ taking the value $s_i=1$ $(-1)$ when
the $i$th particle is of type $A$ $(B)$. The total configurational energy can
then be written

\begin{eqnarray}
\Phi^N(\{r,s\}) &=& \sum_{(i<j)} U_{HS}(r_{ij}) + \sum_{(i<j)} J(r_{ij}) (1+s_i s_j)/2 \nonumber \\
&+& \sum_{(i<j)} \delta J(r_{ij}) (1-s_i s_j)/2 .
\label{MF3}
\end{eqnarray}

To obtain the liquid-vapour phase envelope and the $\lambda$ line, we
adopt an approach similar to that employed in
references~\cite{STRATT1,STRATT2,DESJARDIN,DESMEDT,WILDING2,NIELABA}.
In the thermodynamic limit, the (Helmholtz) free energy density
$f(\rho,T)$ as a function of the number density $\rho=N/V$ and
temperature $T$, is given by:

\begin{equation}
f(\rho,T) = \lim_{V \to \infty} \frac{-1}{\beta V} \ln tr 
\left[ \exp \left( -\beta \Phi^{N} \right) \right] \;\;.
\label{MF2}
\end{equation}
Now, within the mean field approximation, one assumes an interaction between
an A-type particle and an effective field:

\begin{equation}
h_A=J^+ + (J^-/N) \sum_{i>1}s_i = J^+ + J^- m ,
\end{equation}
and an interaction between a B-type particle and an effective field 

\begin{equation}
h_B=J^+ - (J^-/N) \sum_{i>1}s_i = J^+ -J^-m .
\end{equation}
Here, $J^{\pm}$ are effective potentials and $m=(N_A-N_B)/(N_A+N_B)$.
Since the
coordination number in the fluid is density dependent, we make the
approximation:

\begin{equation}
J^{\pm} = -\rho \int d^3r \frac{J^{AA}(r) \pm J^{AB}(r)}{2} g(r),
\label{MF4}
\end{equation}
where the fluid correlation function $g(r)$ is taken from the
Percus-Yevick solution for hard spheres~\cite{HANSEN}.

The mean field configurational energy is then

\begin{eqnarray}
\Phi^N_{MF}(\{r,s\}) &=& U_{HS}(r_{ij}) \nonumber \\
&-& \frac{1}{4}\sum_{i=1}^N \left[h_A(1+s_i)+ h_B(1-s_i)\right] 
\end{eqnarray}
from which the free energy at constant $T$ follows as :

\begin{eqnarray}
f_{MF}(\rho) &=& \lim_{V \to \infty} \frac{-1}{\beta V} \ln tr 
\left[ \exp \left( -\beta \Phi_{MF}^{N}  \right) \right] \nonumber \\
& = & 
f_{HS}(\rho) -\rho \frac{J^+}{2} \nonumber \\
&+& {\rm min}_m \left[ \frac{J^- \rho m^2}{2} - \frac{\rho}{\beta} 
\ln 2 \cosh \left( \beta J^-m  \right) \right].
\label{MF5}
\end{eqnarray}
Here $f_{HS}(\rho)$ is the free energy of a reference system comprising
a hard-sphere single component fluid. The third term of the right hand
side of eqn.~(\ref{MF5}) is minimised for 

\begin{equation}
m = \tanh \left[\beta  J^-m  \right].
\label{MF6}
\end{equation}
As $f(\rho)$ is not always a convex function of the density,  we take
the convex envelope in order to find the coexistence densities for the
first order LV transition. 

We also wish to obtain the phase diagram in the $\mu$--$T$ plane. To
achieve this one needs to consider the grand potential
$f(\rho)+\mu\rho$, with $\mu$ the chemical potential. Minimising this
yields the pressure:

\begin{equation}
p(\rho,\mu) = {\rm min}_{\rho'} \left[ -f(\rho') + \mu \rho' \right] .
\label{MF7}
\end{equation}
The coexistence chemical potential is then found by demanding equality of
both the chemical potential and the pressure in the coexisting phases.

\begin{figure}[hbtp] 
\begin{center}
\begin{picture}(65,65)
\put(0,0){\psfig{file=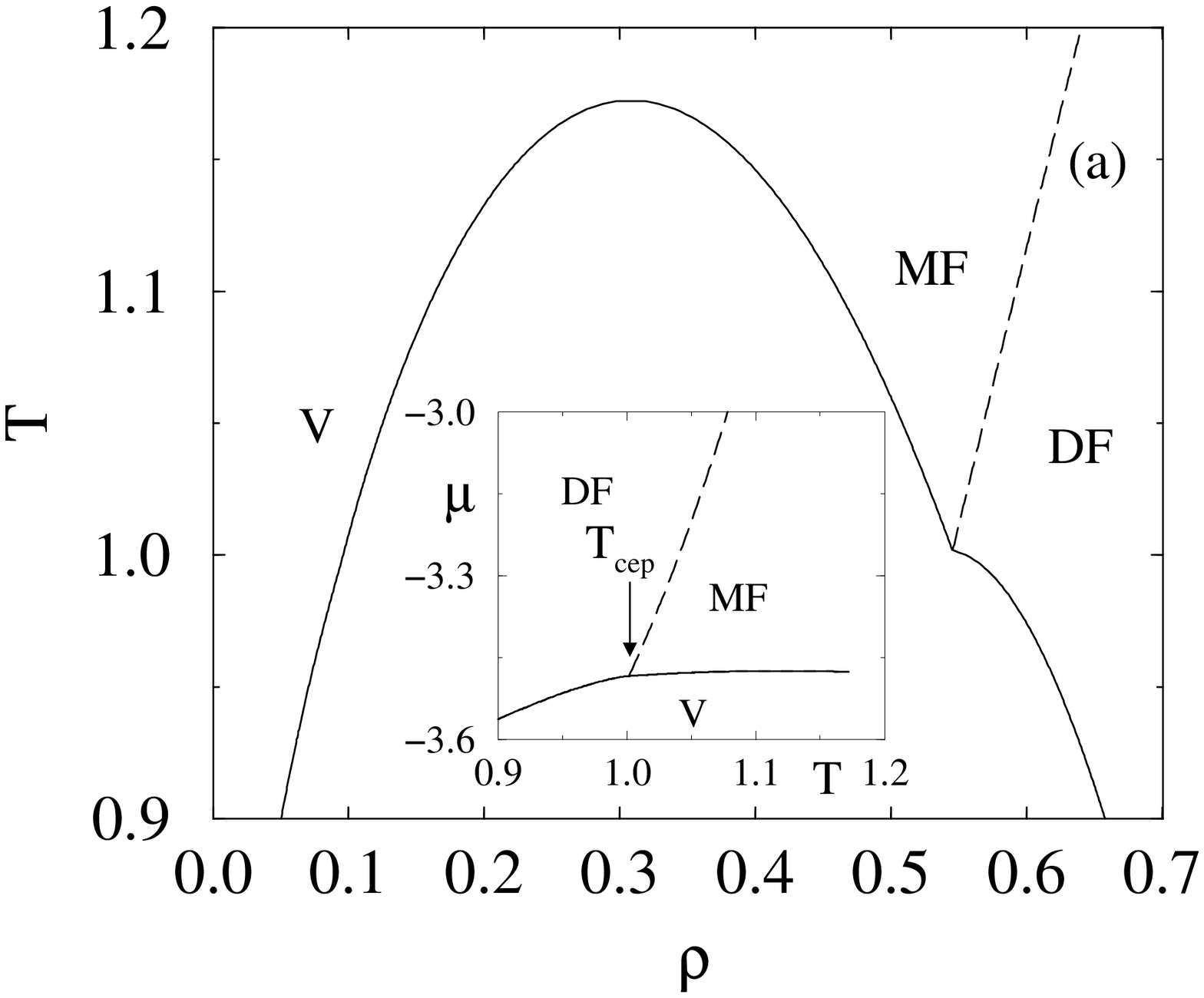,width=70mm,height=65mm}}
\end{picture} 
\begin{picture}(65,65)
\put(0,0){\psfig{file=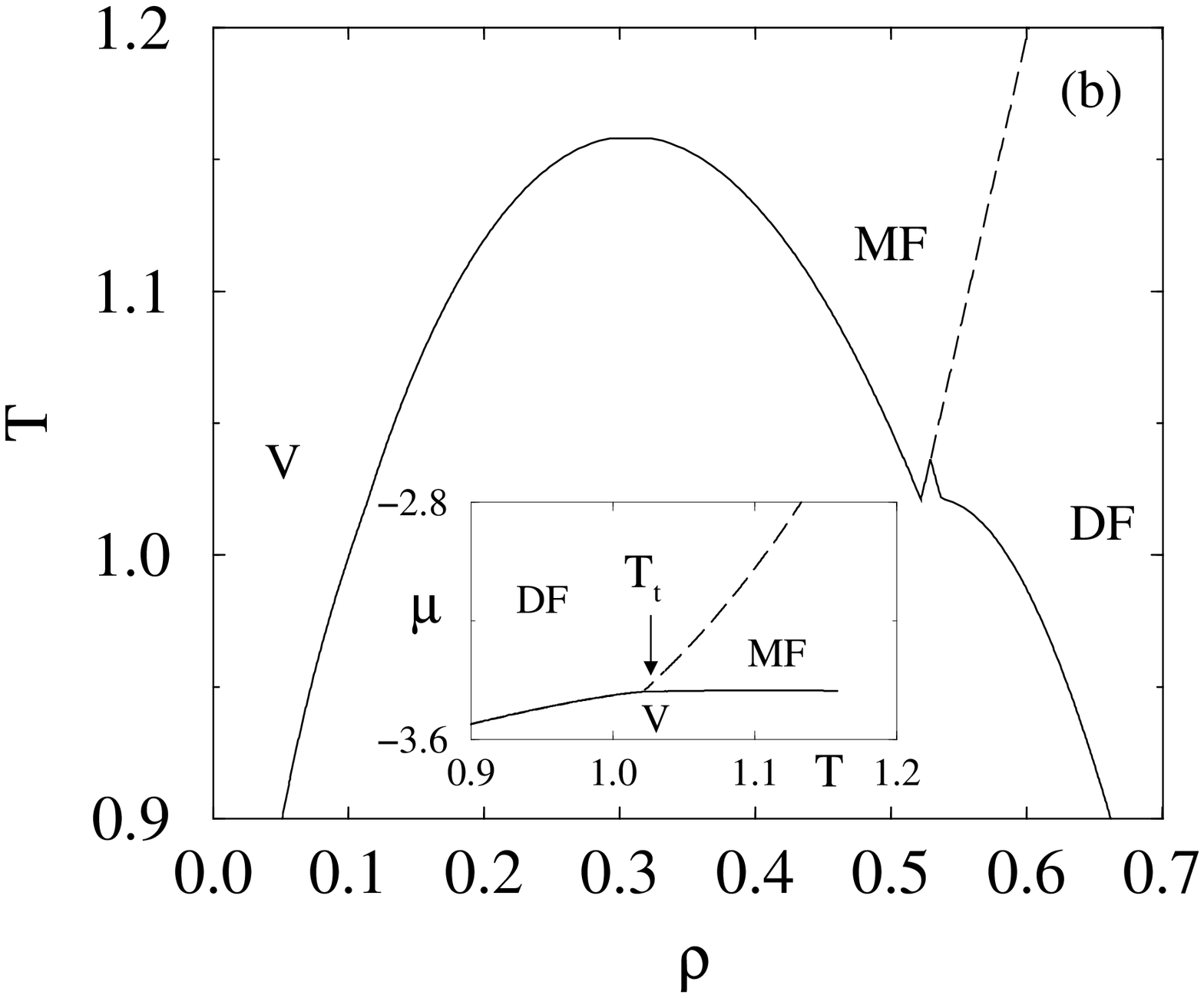,width=70mm,height=65mm}}
\end{picture} \begin{picture}(65,65)
\put(0,0){\psfig{file=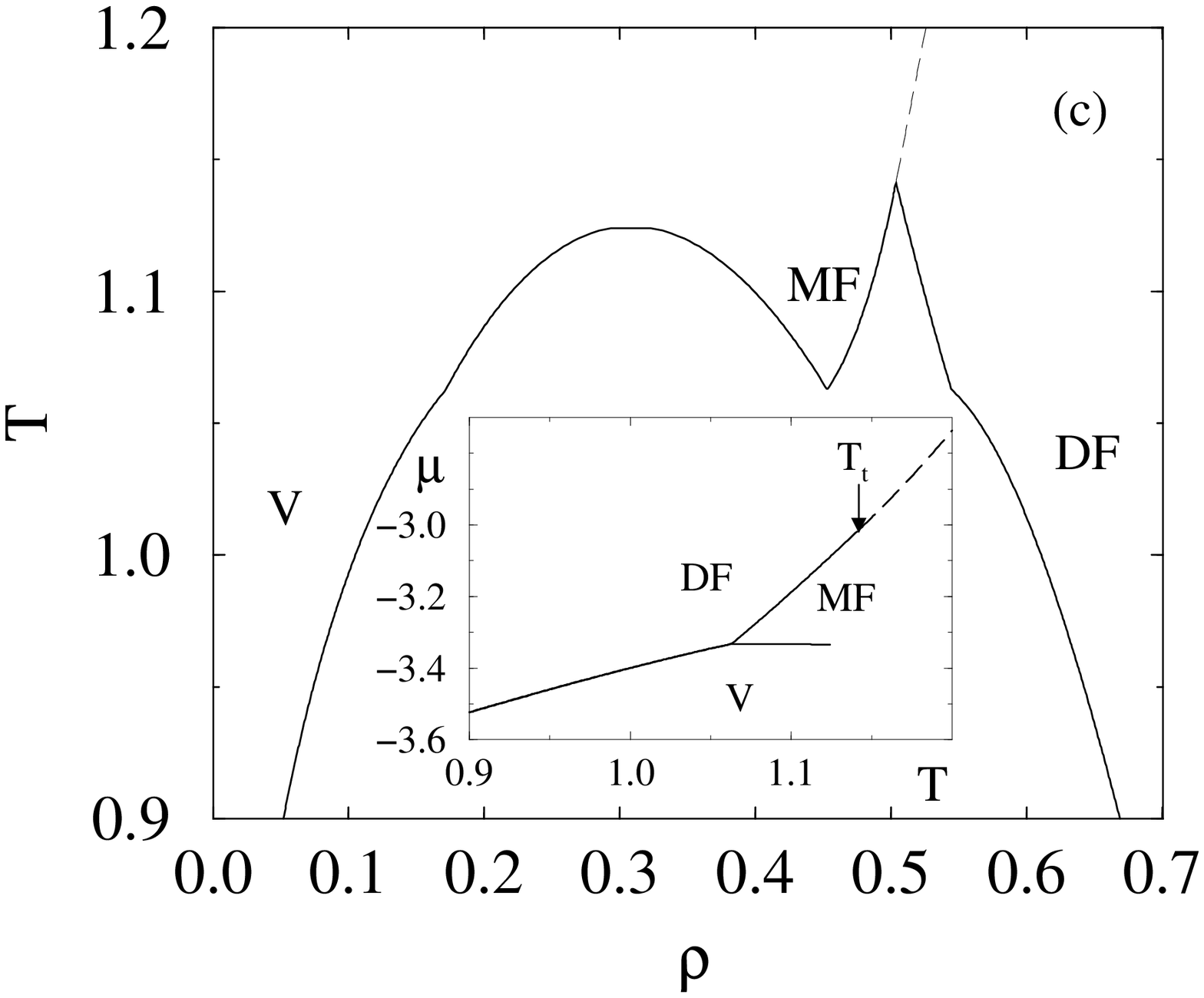,width=70mm,height=65mm}}
\end{picture} 
\newpage
\begin{picture}(65,65)
\put(0,0){\psfig{file=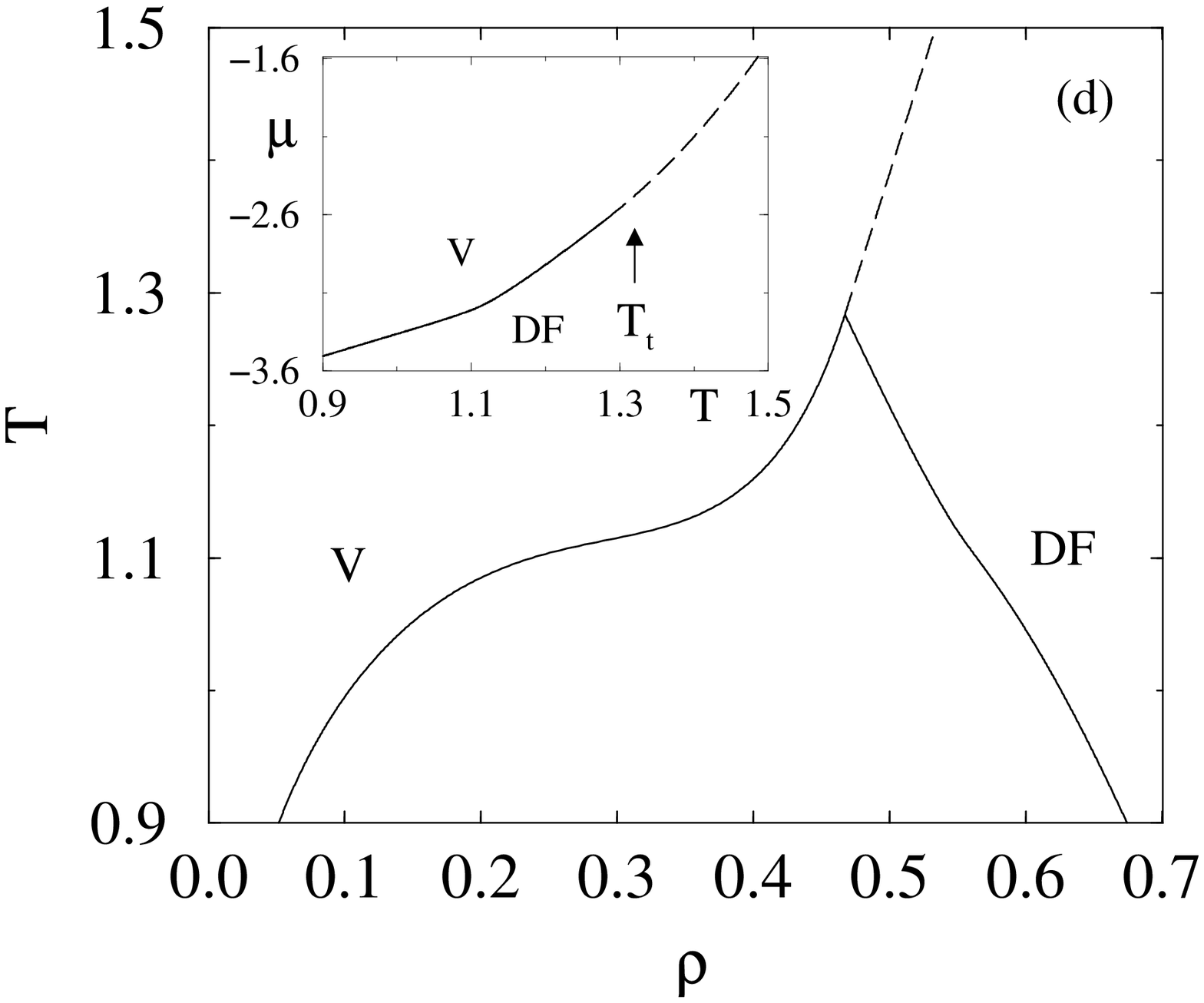,width=70mm,height=65mm}}
\end{picture} 
\vspace*{5mm} 
\caption{Mean field phase diagrams in the $\rho$--$T$ plane for various
$\delta$, as described in the text. {\bf (a)} $\delta=0.72$, {\bf (b)}
$\delta=0.70$, {\bf (c)} $\delta=0.65$, {\bf (d)} $\delta=0.57$. The
insets show the corresponding phase diagrams in the $\mu$--$T$ plane. Full
lines represent first order phase coexistence and dashed curves
represent the $\lambda$ line. The demixed fluid (DF), mixed fluid (MF)
and vapour (V) phases are also marked.}

\label{fig:bmf}
\end{center}
\end{figure}

The resulting phase diagrams in the $\rho$--$T$ plane are shown in
fig.~\ref{fig:bmf}a-\ref{fig:bmf}d, with the $\mu$--$T$ phase
diagrams shown as insets.  For large values of $\delta<1$ we find a LV
coexistence region and a $\lambda$ line at high densities which
intersects the LV line at a critical end point. The CEP induces an
anomaly or kink in the liquid branch density which is clearly visible
in fig.~\ref{fig:bmf}a. This anomaly is the mean field remnant of
the specific heat-like singularity studied theoretically and
computationally in references~\cite{FISHER,WILDING1}. Because the mean
field specific heat exhibits a jump rather than a divergence at
criticality, the anomaly takes the form of an abrupt change in the
gradient of the liquid branch $d\rho_l/dT$. If fluctuations are taken
into account, however, $d\rho_l/dT$ diverges at the CEP
\cite{WILDING1}. 

As $\delta$ is reduced, the CEP anomaly grows until at around
$\delta=0.7$ a small peak emerges [cf. fig~\ref{fig:bmf}b]. The point
at which this occurs constitutes a {\em tricritical end point} as
discussed in subsection~\ref{sec:landau}. On further reduction of
$\delta$, the peak develops until the situation shown in
fig.~\ref{fig:bmf}c is attained. Here the first order coexistence
envelope displays a triple point at which a vapour coexists with a
mixed liquid at intermediate density and a demixed liquid of higher
density. Above the triple point temperature, a vapour and a
demixed liquid coexist at low and moderate densities, merging at the
liquid-vapour critical point. At higher densities a mixed liquid and
the two symmetrical demixed liquids coexist, becoming identical at a
tricritical point.

If $\delta$ is reduced further still, one reaches a point (for $\delta
< 0.605$), at which the triple point temperature equals the LV
critical point temperature. Thereafter, the liquid-vapour critical
point is lost and only a tricritical point remains,  as shown for the
case $\delta=0.57$ in fig.~\ref{fig:bmf}d. No further topological
changes in the phase diagram are observed as $\delta$ is reduced to
zero. 

\unitlength1mm 

\begin{figure}[h]
%\vspace*{5mm}
\setlength{\epsfxsize}{7cm}
\centerline{\mbox{\epsffile{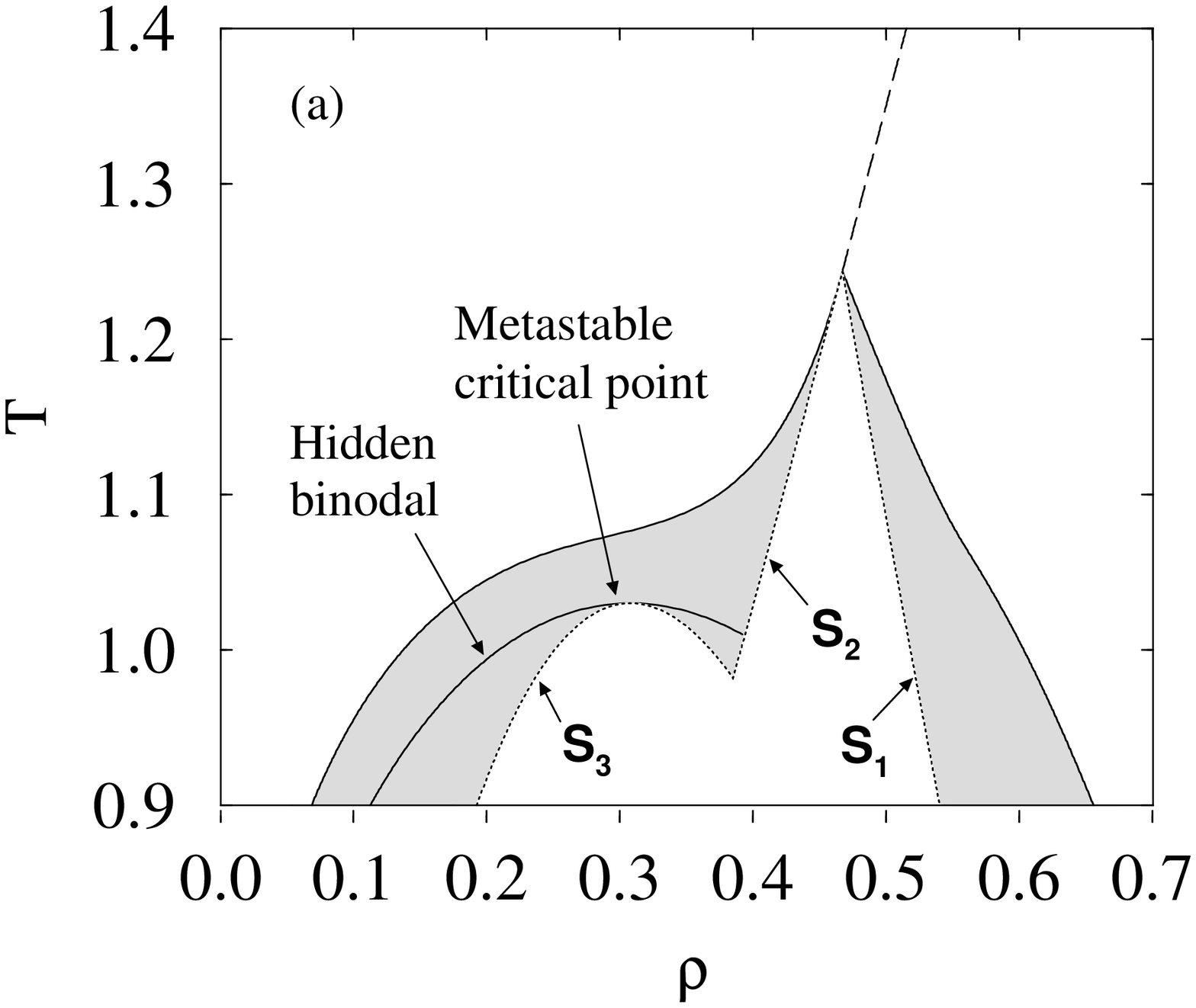}}} 
\vspace*{5mm}
\centerline{\mbox{\epsffile{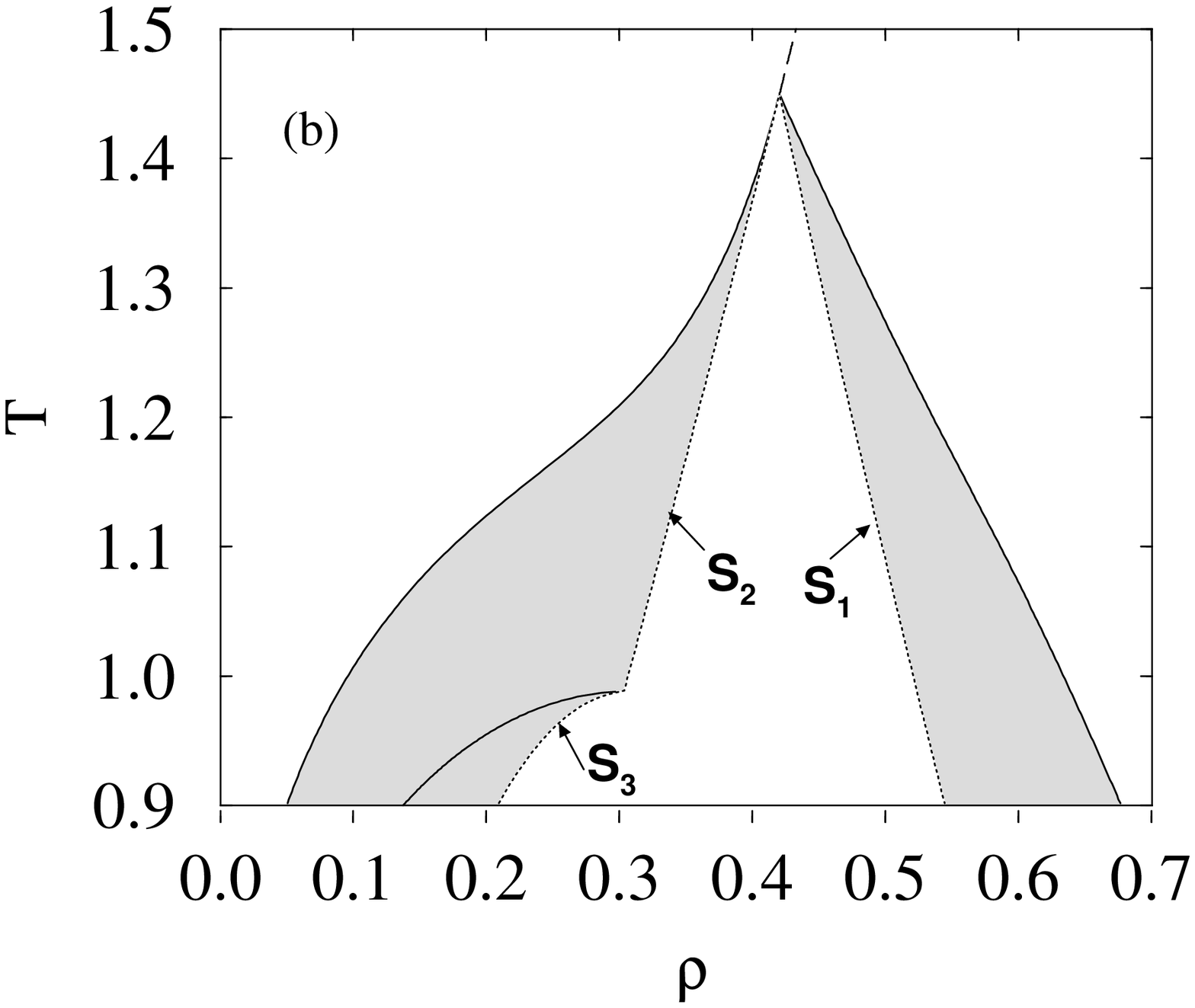}}} 
\vspace*{8mm}

\caption{{\bf (a)} Liquid-vapour phase envelope in the $\rho$--$T$ plane
for $\delta=0.57$. Also shown are the spinodals $S_1, S_2$ and $S_3$
and the `hidden' binodal as described in subsections~\ref{sec:mfexp}
and \ref{sec:landau}. {\bf(b)} The phase envelope and spinodals for
$\delta=0.45$, by which point the metastable LV critical point has been
lost.}

\label{fig:hidden}
\end{figure}

Although the liquid-vapour critical point disappears from the
equilibrium phase diagram for $\delta<0.605$, it is interesting to note
that it nevertheless remains in {\em metastable} form for some range of
$\delta$. This is illustrated in fig.~\ref{fig:hidden}a which shows
the phase diagram for $\delta=0.57$ together with the three spinodals
delineating the limits of metastability of the demixed fluid, mixed
fluid and vapour (marked $S_1$, $S_2$ and $S_3$). Also shown is the
`hidden binodal' for coexistence between a vapour and a demixed liquid,
calculated by neglecting the coupling of the concentration to the
density. Clearly for this value of $\delta$, the hidden binodal and the
liquid-vapour critical point (at which it terminates) both lie within
the metastable region.  For smaller $\delta\lesssim 0.45$, however, the
metastable critical point moves outwith the limit of metastability and
the hidden binodal is lost [ig.~\ref{fig:hidden}b].  Although
the phases corresponding to the hidden binodal are not observable at
equilibrium, they {\em are} expected to influence the dynamical
properties of the system.  We return to this point in  more detail in
subsection~\ref{sec:landau}.

\subsection{General Landau theory considerations}

\label{sec:landau}

\newcommand{\r}{\varrho}
\newcommand{\hmu}{\hat{\mu}}
\newcommand{\hm}{\hat{m}}

We now turn to analyse the system from a more general point of view.
Clearly we are dealing with a two order parameter problem, {\em i.e.}
the density $\rho=N/V$ and the number difference order parameter
$m=(N_A-N_B)/N$. In a symmetrical fluid, the Hamiltonian has to be
invariant under sign reversal of $m$. The Landau expansion of the grand
potential thus takes the general form 

\begin{eqnarray}
\label{ff} 
F& = & a \frac{(\rho-\rho_0)^2}{2}  + \frac{(\rho-\rho_0)^4}{4}  - \mu (\rho-\rho_0) \\ 
&& + A \frac{m^2}{2} + \frac{m^4}{4} - \frac{B}{2} m^2 (\rho-\rho_0) , \nonumber
\end{eqnarray} 
where $\mu$ is the chemical potential, and $\rho_0$ is a reference
density in the liquid-vapour coexistence region, chosen such that the
cubic term  $\propto (\rho-\rho_0)^3$ vanishes. An expansion of this
type has been discussed in a different context by Roux {\em et al} 
\cite{ROUX}; it applies generally to fluids with an
additional Ising like ordering tendency. In the case of Heisenberg 
type ordering,  e.g., in a ferromagnetic fluid, the Landau
expansion  looks very similar with $m$ simply replaced by the vector
$\vec{m}$, and many of the conclusions drawn below still hold. The
liquid-vapour critical point of the mixed fluid is found at $a=1$,
and could the fluid be kept at fixed density $\rho_0$, it would demix
or order at $A=0$.

Phase diagrams of such fluids are often discussed in terms of an
`interaction ratio' $R$, comparing the strength of the ordering or
demixing tendency in the fluid with the overall attractive interactions
between particles \cite{HEMMER,TAVARES,WEIS}.  $R$ corresponds to
$1-\delta$ in our model, to $J/K$ in the Blume-Emery-Griffiths model
\cite{BEG}, to the dipole moment $m$ in reduced  units in \cite{GROH}
and to $1/T_c^0$ in Reference \cite{ZHANG}.  It is generally found that
increasing $R$ drives the phase behaviour from the topology depicted in
fig~\ref{fig:schem}a via ~\ref{fig:schem}b
towards~\ref{fig:schem}c, {\em i.e.} a critical end point turns into a
tricritical point, which moves up in temperature, until the
liquid-vapour critical point disappears in the region of coexistence
between demixed and mixed liquid phases.

Although the interaction ratio appears to be an influential quantity,
it is important to note that the basic factor driving the phase
behaviour  is the {\em coupling} between the two order parameters. In
eq (\ref{ff}),  it is described by the last term. At $B$=0, no
tricritical point can be expected, regardless of the interaction ratio.
Therefore, in what follows we shall analyse the phase behaviour in terms
of the coupling strength rather than the interaction ratio. The
relation between the two will be discussed later.

We will focus on the transition from a topology with a critical end point to 
one with a tricritical point, and assume that we are well below the 
liquid-vapour critical temperature, $a < 0$. It is then convenient
to rescale the Landau expansion eqn (\ref{ff}) such that
\begin{equation}
\label{ff2}
F = \theta \frac{\hm^2}{2} + \frac{\hm^4}{4} 
- \frac{\r^2}{2} + \frac{\r^4}{4} - \hmu \r + \kappa (1-\r) \hm^2,
\end{equation}
where $\hmu=\mu/(\sqrt{-a})^3$, $\r=(\rho-\rho_0)/\sqrt{-a}$, 
$\hm=m/\sqrt{-a}$, 
and $F$ is written in units of $a^2$. The parameter $\kappa = B/2\sqrt{-a}$ 
then describes the effective coupling between the order parameter and the
density. The parameter $\theta=A/(-a) - 2 \kappa$ is temperature-like: 
$\theta \propto (T-T_{CEP})$, where $T_{CEP}$ is the temperature of the
(stable or unstable) critical demixing point at fixed density $\r = 1.$
The phase behaviour is found by minimising $F$ with respect to $\hm$ and $\r$.

At $\kappa = 0$, the ordering behaviour and the liquid-vapour phase
separation decouple and do not affect one another. One finds
coexistence of a liquid ($\r=1$) and a vapour ($\r=-1$) at $\hmu=0.$, and
these phase boundaries are crossed by the critical ordering ($\lambda$)
line at $\theta=0.$ If a small coupling $\kappa$ is turned on, the
ordering temperature increases with the density:

\begin{equation}
\label{cl}
\theta_c(\r) = 2 \kappa (\r-1).
\end{equation}
Thus the $\lambda$ line shifts at the LV coexistence line. Technically,
the Landau expansion (\ref{ff2}) predicts two critical end points, one
on the liquid side at $\theta = 0$, and one on the vapour side at
$\theta = -4 \kappa$. In all fluids studied so far, only the upper one
has been seen. Situations with two CEPs are however  encountered when a
critical line intersects a liquid-solid coexistence region \cite{HEMMER,GROH}.

Next we study the stability of the demixed liquid phase. The order parameter in the 
homogeneous demixed liquid takes the value $\hm^2 = \theta_c(\r) - \theta$.
The determinant of the stability matrix there is given by
$4 \hm^2((3 \r^2-1)/2-\kappa^2)$. Hence the demixed phase becomes unstable 
for $\r < \r_c$, with the spinodal line
\begin{equation}
\r_c = \sqrt{(2 \kappa^2+1)/3}.
\label{eq:sl}
\end{equation}
A tricritical point is found when the spinodal line intersects the critical
line (\ref{cl}). This requires $\r_c > 1$, {\em i.e.} the coupling 
$\kappa$ has to be larger than a limiting value $\kappa_0 = 1$.
The tricritical point is then located at
\begin{eqnarray}
\theta_t = \theta_c(\r_c) \quad && \mbox{and} \quad \r_t = \r_c  \\  
\mbox{or} && \quad
\hmu_t =  \frac{2}{3} \sqrt{\frac{1+2 \kappa^2}{3}} \; (\kappa^2-1).
\end{eqnarray}

We conclude that the coupling $\kappa$ between the order parameter and
the density determines the topology of the phase diagram. If $\kappa$
exceeds $\kappa_0$, the topology switches from one with a critical end
point [fig.~\ref{fig:schem}a] to one with a tricritical point
[fig.~\ref{fig:schem}b]. From a physical point of view, $\kappa$
basically reflects the correlations  between order parameter and
density. For example, the response of the average  order parameter to a
change of chemical potential in the demixed liquid phase is given by

\begin{equation} \frac{\partial \hm}{\partial \hmu} \propto
\langle \hm \r \rangle - \langle \hm \rangle \langle \r \rangle
\propto \kappa \frac{(\theta_c-\theta)^{-\zeta}}{\r - \r_c} ,
\end{equation}
where the exponent $\zeta$ is given by $\zeta = 1/2$ in mean field theory.
(Scaling arguments yield $\zeta = (\gamma + \alpha)/2$, where
$\gamma$ and $\alpha$ are the usual Ising critical exponents of the 
order parameter susceptibility and the specific heat).

It is instructive to investigate the relationship between $\kappa$ and
the interaction ratio $R$. When $R$ increases, the critical end point
temperature $T_{CEP}$ moves  closer to the liquid-vapour critical
temperature $T_{c0}$. Assuming  $a \propto (T_{c0}-T)$, this implies
that the effective coupling $\kappa \propto \sqrt{1/(T_{c0}-T)}$
increases also, and diverges as $T_{CEP}$ approaches $T_{c0}$. As long
as there is any coupling between the density and the order parameter
({\em i.e.} $B>0$ in eq.(\ref{ff})),  tuning $R$ has the effect of tuning the
coupling. Thus our arguments are supported by our own explicit model
calculations and by the results  quoted earlier
\cite{HEMMER,TAVARES,WEIS,ZHANG,GROH,BEG}.

Next we discuss the implications for the LV phase boundary.  The small
coupling limit $\kappa < \kappa_0$ has been studied in detail in ref.
\cite{WILDING1}. In chemical potential space, the critical end point
induces a weak singularity in the first order liquid-vapour line
\cite{FISHER},  

\begin{equation}
\mu(T) = \mu_{reg}(T) - U|t|^{2-\alpha},
\end{equation}
where $\mu_{reg}(T)$ is an analytical function of the temperature $T$,
$t=(T-T_{CEP})/T_{CEP}$ and $U$ is a critical amplitude. In the Landau
theory framework, $\mu_{reg}$ is simply given by
$\mu_{reg}=\hmu_{reg}\equiv 0$. 

The density of the vapour phase at coexistence behaves in a similar way

\begin{equation}
\rho_g(T) = \rho_{g,reg}(T) - V_g|t|^{2-\alpha},
\end{equation}  
whereas the density of the liquid phase shows the marked hump indicated
in fig.~\ref{fig:schem}a, given by \cite{WILDING1}
\begin{equation}
\rho_l(T) = \rho_{l,reg}(T) - V_l|t|^{1-\alpha}.
\label{eq:liqsing}
\end{equation}

In the strong coupling limit, $\kappa > \kappa_0$, one deals with the
simple case in which two first order lines meet at a triple point. One
thus expects a kink in $\mu(T)$ and in $\rho_g(T)$, and the density of
$\rho_l(T)$ jumps to the coexisting demixed liquid phase [cf.
fig.~\ref{fig:bmf}c]. The two regimes meet at $\kappa = \kappa_0$,
where the critical end point turns into a tricritical end point (TEP)
\cite{TEPNOTE} (cf. fig~\ref{fig:bmf}b). Within our Landau theory, we
have  calculated the phase boundaries in the vicinity of a tricritical
end point. Since we expect that $\kappa$ varies with temperature, we
consider a path  $\kappa = \kappa_0 + K \theta$, where $\theta \propto
(T-T_{TEP})$ as defined above. Our results to leading order in
$\theta$ are summarised as follows: For the chemical potential of the
liquid vapour line, we find

\begin{equation}
\hmu = 0 \quad (\theta > 0), \qquad 
\hmu = - (|\theta|/3)^{3/2} \quad (\theta < 0).
\end{equation}
The rescaled densities of the vapour phase (spectator phase, $\r_-$) and the
liquid phase ($\r_+$) are given by
\begin{displaymath}
\r_{\mp} = \mp 1 \qquad (\theta > 0) 
\end{displaymath}
\begin{equation}
\begin{array}{ccccc}
\r_- &=& -1 &-& 1/2 \; (|\theta|/3)^{3/2}  \\
\r_+ &=& 1 &+& (|\theta|/3)^{1/2}  
\end{array}
\qquad (\theta < 0).
\end{equation}
Note that these results agree with the arguments presented in
\cite{WILDING1}  if one inserts $\alpha_t = 1/2$, the mean field value
of the exponent  $\alpha$ at the tricritical point. Mean field theory
is expected to yield the correct tricritical behaviour in our system,
since the upper critical dimension of a tricritical point  is $d_u=3$.

We close this subsection with a  discussion of `hidden' parts of the phase
diagram. To this end, we return to eqn~\ref{ff}  and perform a general
stability analysis. We consider the demixed phase at
$m=[A-B(\rho-\rho_0)]^{1/2}$ and the mixed phase at $m=0$. 

%The stability matrix is then given by
%\begin{equation}
%\label{d2fo}
%\partial^2 F = \left( 
%\begin{array}{cc} a+3 (\rho-\rho_0)^2 & -B m \\-B m &2 m^2 \end{array}
%\right)
%\end{equation}
%in the ordered phase, and
%\begin{equation}
%\label{d2fd}
%\partial^2 F = \left( 
%\begin{array}{cc} a+3 (\rho-\rho_0)^2 & 0\\0&A-B (\rho-\rho_0) \end{array}
%\right)
%\end{equation}
%in the disordered phase. 

The conditions for stability of the demixed phase have already been
discussed earlier. One finds a spinodal at 

\begin{equation} 
\rho-\rho_0 = \pm S_1 \quad \mbox{with} \quad S_1^* = \sqrt{(B^2-2 a)/6},
\end{equation} 
which is equivalent to eqn.~\ref{eq:sl} in non rescaled units.  For
$|\rho-\rho_0| < S_1^*$, the demixed phase becomes unstable with
respect to phase separation into a demixed liquid and a vapour.

In the present context, the stability of the mixed liquid phase is more 
interesting. The stability analysis yields two spinodals: At

\begin{equation}
|\rho-\rho_0| < \pm S_2^* \quad \mbox{,} \quad S_2^* = \sqrt{-a/3},
\end{equation}
the demixed liquid phase is unstable with respect to phase separation into
a demixed liquid and its vapor, and at 

\begin{equation}
\rho-\rho_0 > S_3^* = A/B,
\end{equation}
it becomes unstable with respect to demixing. As long as $A>0$ and the coupling
$B$ is sufficiently small, there exists a region on the high density
side of  the fluid, $\rho>\rho_0$, where the mixed liquid phase can be
metastable or  even stable 

\begin{displaymath}
\sqrt{-a/3} < \rho-\rho_0 < A/B
\end{displaymath}
In that case, a binodal can be found at $|\rho-\rho_0| = \sqrt{-a}$, 
which may be stable or hidden in the metastable region. 

The spinodals and the hidden binodal are indicated in fig.~\ref{fig:hidden}. Note
that both the coefficients $a$ and $A$ depend roughly linearly on the
temperature $T$. This explains the linear form of $S_2$ in the 
density-temperature plane as opposed to the parabolic form of $S_3$.

The hidden binodal disappears completely as soon as $A\le 0$ at $a=0$, 
{\em i.e.}, as soon as the spinodal $S_3$ meets $S_2$ at the `hidden
critical point' or beyond (at $\rho < \rho_0$). This criterion is 
independent of the coupling $B$. The strength of coupling thus has no 
influence on the appearance of a hidden binodal. It does however affect
the range of metastability of the hidden binodal, {\em i.e.}, the
temperature interval before it is lost by intersecting the spinodal
$S_2$. 

Of course, the concepts of spinodals and metastability only really make
sense within a mean field treatment, and strictly speaking lose their
physical meaning as soon as fluctuations are taken into account.
However, a discussion of the metastable and unstable regions is still
useful in the context of the dynamical properties of the system.
Consider, for instance, a binary fluid with interactions corresponding
to the situation shown in fig.~\ref{fig:hidden}a, at density
$\rho=\rho_0$, which is quenched from some high temperature into the 
coexistence region slightly below the `metastable critical point'. One can
then expect `two-stage demixing'. In the first stage, the fluid will
separate into domains of vapour and mixed liquid, and the separation of
these domains will be accelerated by the driving force of gravitation.
In the second stage, the liquid phase will slowly demix, and droplets
of demixed liquid will additionally nucleate from the vapour phase. If the
interactions of the fluid correspond to the situation depicted in
fig.~\ref{fig:hidden}b, on the other hand, no intermediate stage
will appear and the fluid will demix and phase separate simultaneously.

\section{Monte-Carlo studies}
\label{sec:mc}

\subsection{Simulation details}
\label{sec:simn}

Many features of the simulation techniques employed in the present
study have previously been detailed elsewhere
\cite{WILDING0,WILDING3,WILDING2}. Accordingly, we confine the
description of our methodology to its barest essentials, except where
necessary to detail a new aspect.

We assume our system to be contained in volume $L^3$ and to be
thermodynamically open so that the total number density and
concentration can fluctuate. The associated (grand canonical) partition
function is:

\begin{equation}
\label{eq:bigzdef}
{\cal Z}_L  = \sum _{N=0}^{\infty }\sum_{\{s_i\}}\prod _{i=1}^{N}
\left\{\int d\vec{r}_i\right\} e^{-\beta\left[\Phi (\{ \vec{r},s\} )
+ \mu N \right] }
\label{eq:pf}
\end{equation}
Here,  the species label $s_i=1,-1$ denotes respectively the two
particle species $A$ and $B$, $N=N_A+N_B$ is the total particle number,
 $\beta=1/k_BT$ is  the inverse temperature and $\mu$ is the chemical
potential. The configurational energy density $\Phi$ is given by

\begin{equation}
\Phi (\{ \vec{r},s \} )=\sum_{ij}U(r_{ij},s_is_j)
\end{equation}
where the symmetrical square-well interparticle potential $U$ is defined in
eqs.~\ref{eq:potdef} and~\ref{eq:symmdef}.

Grand canonical Monte Carlo (MC) simulations were performed using a
standard Metropolis algorithm \cite{ALLEN,WILDING3}. The MC scheme
comprises two types of operations:

\begin{enumerate}
\item Particle insertions and deletions. 

\item Particle identity transformations: $A\to B$, $B\to A$
\end{enumerate}
Since particle positions are sampled implicitly via the random particle
transfer step,  no additional translation algorithm is required.

To simplify identification of particle interactions, we employed a
linked list scheme \cite{ALLEN}. This involves partitioning the
periodic simulation space of volume $L^3$ into $l^3$ cubic cells, each
of linear dimension the interaction range, {\em i.e.} $L/l=1.5$.
We chose to study two system sizes corresponding to $l=8$ and $l=10$,
containing, at LV coexistence, approximate average particle numbers
$\langle N \rangle=550$ and $1100$ respectively. Equilibration periods
of up to $2\times10^6$ MCS were employed and sampling frequencies were
$100$ MCS for the $l=8$ system to $150$ MCS for the $l=10$ system.
Production runs amounted to $2\times10^7$ MCS for the $l=8$ and
$5\times10^7$ MCS for the $l=10$ system size. At coexistence, the
average acceptance rate for particle transfers was approximately
$10\%$, while for spin flip attempts the acceptance rate was
approximately $40\%$. 

In this work we wish to explore the parameter space spanned by the
three variables ($\mu,T,\delta)$. To accomplish this, without having to
perform a very large number of simulations, we employed the histogram
reweighting technique \cite{FERRENBERG}. Use of this technique permits
histograms obtained at one set of model parameters to be reweighted to
yield estimates appropriate to another set of model parameters. To
enable simultaneous reweighting in all three fields $\mu,T,\delta$, one
must sample the conjugate observables $(\rho,u,u_d)$, with $\rho=N/V$
the number density,  $u=\Phi/V$ the configurational energy density, and
$u_d$ that part of $u$ associated with interactions between {\em
dissimilar} particle species. In addition to these variables, we have
also accumulated the quantity $\tilde{m}=(N_A-N_B)/V=\rho m$ which
gives a measure of the degree of $A-B$ ordering in the system.

As mentioned in the introduction, standard GCE simulations, deep within
the LV coexistence region, are hampered by the large free energy
barrier separating the two coexisting phases. This barrier leads to
metastability effects and prohibitively long correlation times.  To
circumvent this difficulty, we have employed the multicanonical
preweighting method \cite{BERG} which encourages the simulation to
sample the interfacial configurations of intrinsically low probability.
This is achieved by incorporating a suitably chosen weight function in
the MC update probabilities. The weights are subsequently `folded out'
from the sampled distribution to yield the correct Boltzmann
distributed quantities. Use of this method permits the direct
measurement of the distribution of observables at first order phase
transitions, even when these distributions span many decades of
probability. Details concerning the implementation of the techniques
can be found in references \cite{BERG,WILDING2}.

\subsection{Method and results}

Using the multicanonical simulation scheme, we have obtained the
density distribution $p(\rho)$ for a number of states close to the LV
coexistence curve, and for a number of choices of $\delta$. 
We begin, by probing the regime of critical end point behaviour. 

\begin{figure}[h]
\vspace*{0.1 in}
\setlength{\epsfxsize}{7cm}
\centerline{\mbox{\epsffile{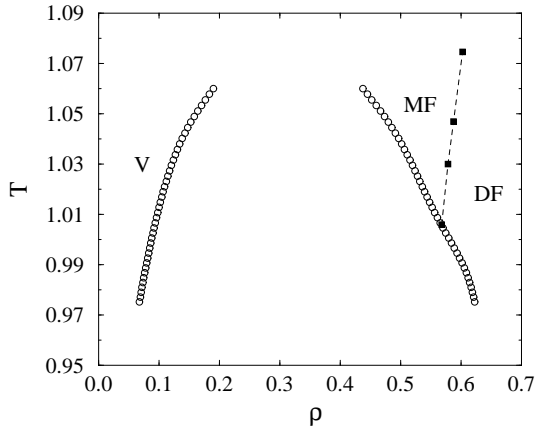}}} 

\caption{The liquid-vapour coexistence curve in the $\rho$--$T$ plane
for $\delta=0.72$, showing the vapour (V), mixed fluid (MF) and demixed
fluid (DF) phases. The results were obtained from the measured peak
positions of the coexistence density distributions for the $l=8$ system
size. Statistical errors do not exceed the symbol sizes.}
\label{fig:simd0.72} \end{figure}

On the basis of the mean field results of sec.~\ref{sec:mf}, CEP
behaviour is expected to occur for large $\delta<1$.  For
$\delta\lesssim 1$, the CEP will occur at very low temperatures
relative to the LV critical point ({\em i.e.} $T\ll T_{c0}$),  but will
move to higher temperatures as $\delta$ is reduced. At some point as
$\delta$ is reduced, the phase diagram is predicted to evolve into a
triple point topology. Thus in seeking to observe CEP behaviour one
should aim to set $\delta$ large and to search at low temperatures.
Unfortunately, very low temperatures are associated with high
liquid densities at LV coexistence, and these inaccessible to our
GCE scheme due to the prohibitively small particle transfer acceptance
rate. In fact we find that the largest value of $\delta$ for which the
density fell within the accessible range ($\rho\lesssim 0.7$) was
$\delta=0.72$.  Although this value of $\delta$ is not as large as one
might hope to attain, it nevertheless transpires that CEP behaviour
occurs. This is demonstrated in fig.~\ref{fig:simd0.72} which shows
the liquid and vapour coexistence densities for this value of $\delta$
obtained as the first moment of the respective peaks of $p(\rho)$ for
the $l=8$ system size \cite{WILDING1}. The data were obtained by
reweighting \cite{FERRENBERG} four histograms spanning temperatures in
the range $T=0.99-1.05$, and coexistence was located using the equal
peak-weight criterion for $p(\rho$)\cite{EWC,WILDING3}. Also shown in
the figure is the measured locus of the critical line.

Clearly the data of fig.~\ref{fig:simd0.72} display an anomaly in the
liquid branch density close to the intersection point of the $\lambda$
line and the liquid branch ({\em i.e.} at the CEP). In the
thermodynamic limit, the liquid branch density is expected to exhibit a
cusp-like singularity at the CEP, as given in
equation~\ref{eq:liqsing}. In our finite-sized system, however, this
critical singularity is smeared out and shifted, so that only a rounded
depression in the coexistence envelope is visible \cite{WILDING1}.
Since these aspects of the CEP singularities have recently been
discussed in detail elsewhere~\cite{WILDING1}, we shall not pursue them
further here. Instead we shall proceed to consider what happens as
$\delta$ is made smaller still.

\begin{figure}[h]
\vspace*{0.1 in}
\setlength{\epsfxsize}{7cm}
\centerline{\mbox{\epsffile{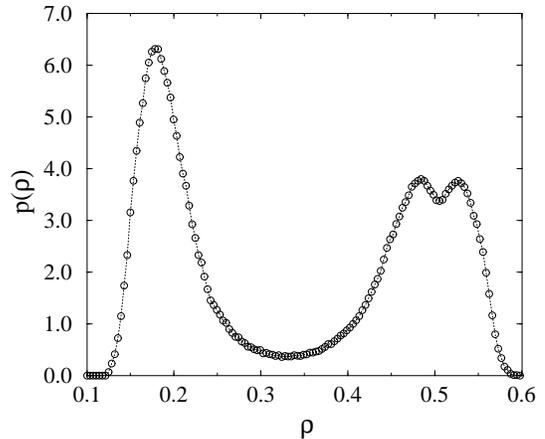}}} 

\caption{The measured near-coexistence density distribution $p(\rho)$
for $T=1.044,\delta=0.675$ showing the three-peak structure discussed
in the text. The distribution is normalised to unit integrated weight
and statistical errors are comparable with the symbol sizes.}

\label{fig:TEP}
\end{figure}

Further reducing $\delta$ continues to shift the CEP closer to the LV
critical point. As one reaches $\delta=0.675$, however, the phase
diagram changes topology. We find that above a certain temperature the
liquid peak in $p(\rho$) decomposes into two peaks. This is shown in
fig.~\ref{fig:TEP} for the $l=10$ system size at a temperature
$T=1.044$. Evident from this figure are two closely separated
overlapping peaks, the presence of which signifies incipient triple
point behaviour. It follows that for this $\delta$ and $T$, the system
lies close to the tricritical end point which heralds entry into the
triple point phase diagram topology [cf. fig.~\ref{fig:bmf}b].
Actually we believe the TEP lies close to $\delta=0.68$ since this is
the value at which we first observe the appearance of a shoulder in the
liquid peak. In a sufficiently large system, this shoulder would
presumably resolve itself into a distinct peak. We have not, however,
attempted to pinpoint the location of the TEP more precisely, as this
would require a full finite-size scaling analysis--a task beyond the
scope of the present study.

\begin{figure}[h]
\vspace*{0.1 in}
\setlength{\epsfxsize}{7cm}
\centerline{\mbox{\epsffile{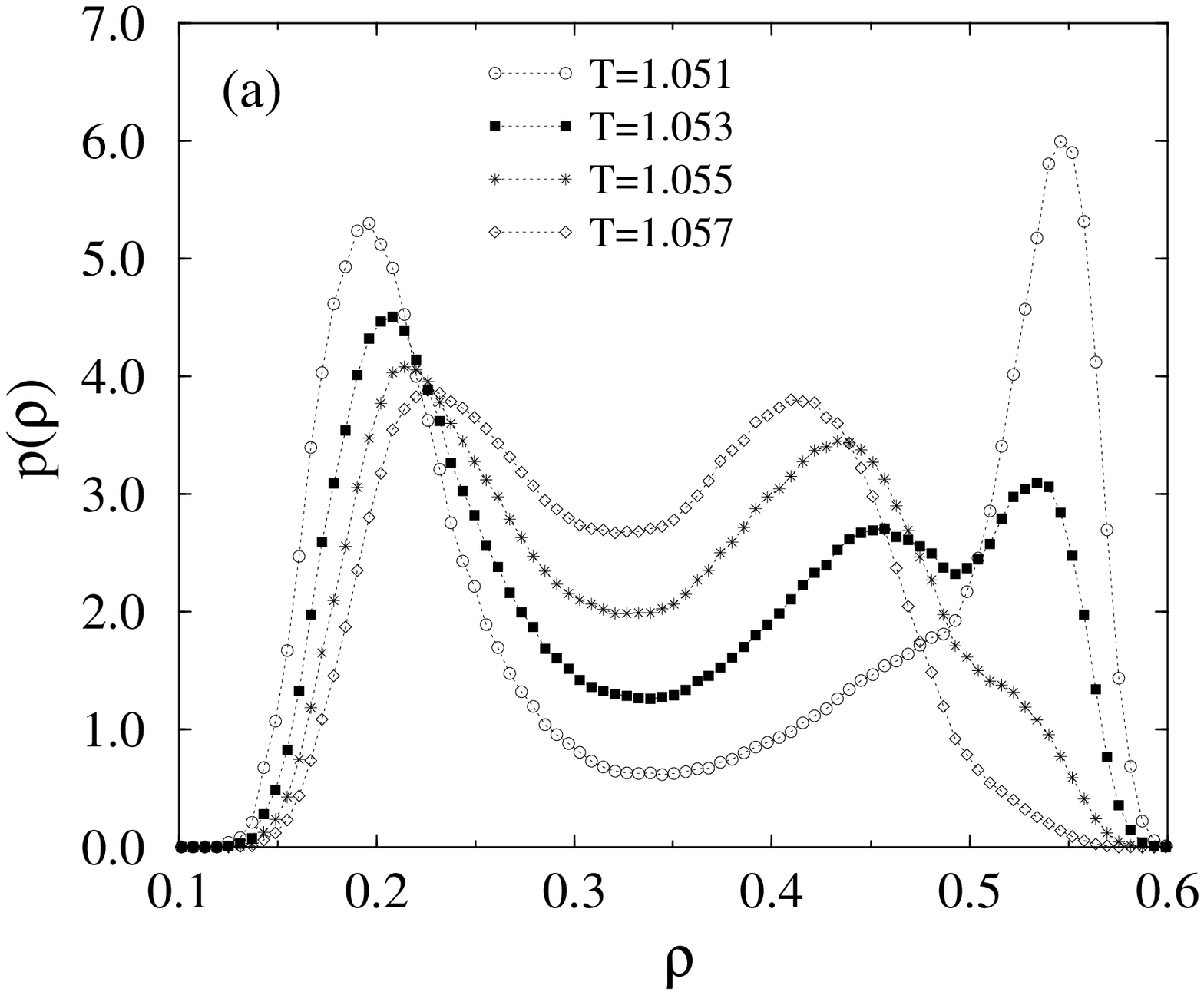}}} 
\centerline{\mbox{\epsffile{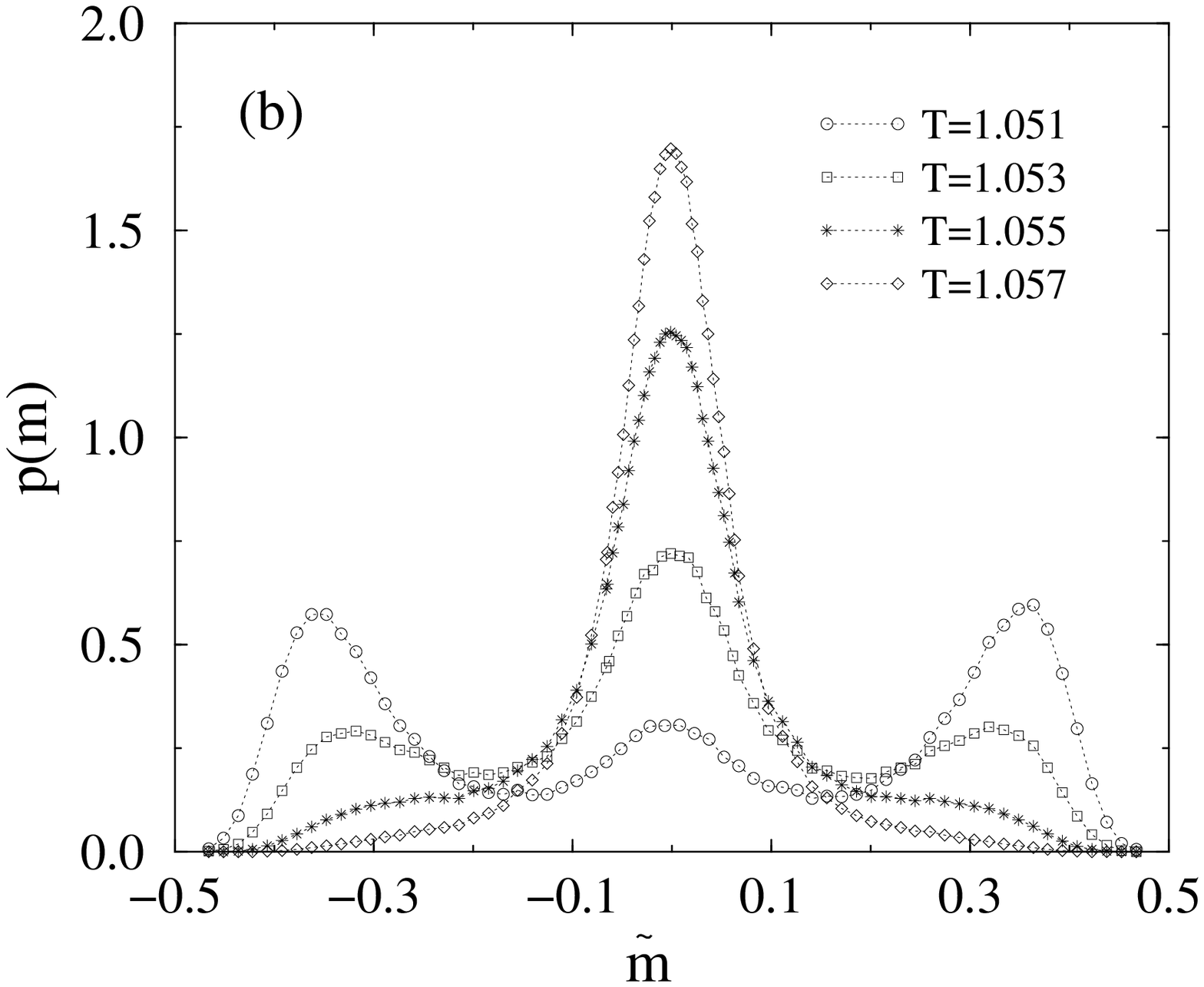}}} 

\caption{{\bf (a)} Coexistence density distributions $p(\rho)$ for
$\delta=0.665$ at a selection of temperatures spanning the triple point
temperature. {\bf (b)} The corresponding form of $p(\tilde{m})$ where
$\tilde{m}=m\rho=(N_A-N_B)/V$. Lines are guides to the eye. Statistical
errors are comparable with the symbol sizes.}

\label{fig:triple}
\end{figure}

Using histogram reweighting, we have monitored the temperature
dependence of $p(\rho)$ as $\delta$ is reduced below the value at which
the TEP occurs. fig.~\ref{fig:triple}a shows a selection of density
distributions for $\delta=0.665$, for which a triple point (TP) occurs
at some temperature $T_{TP}<T_{c0}$, {\em i.e.} below the liquid-vapour
critical temperature. The corresponding forms of $p(\tilde{m})$ are shown in
fig.~\ref{fig:triple}b. At the triple point, a demixed liquid  coexists
with a mixed liquid and its vapour. For $T>T_{TP}$, there is phase
coexistence either between the mixed liquid and its vapour, or between
the mixed and demixed liquid [cf. fig.~\ref{fig:bmf}c]. The
liquid-vapour coexistence terminates at the LV critical point, while
the mixed-demixed liquid coexistence curve terminates at a tricritical
point. From fig.~\ref{fig:triple}a one sees that for $\delta=0.665$,
the tricritical point temperatures lies slightly below the LV critical
point temperature, as evidenced by the fact that on increasing $T$ the
liquid peaks merge before the liquid and vapour peaks do so.

\begin{figure}[h]
\vspace*{0.1 in}
\setlength{\epsfxsize}{7cm}
\centerline{\mbox{\epsffile{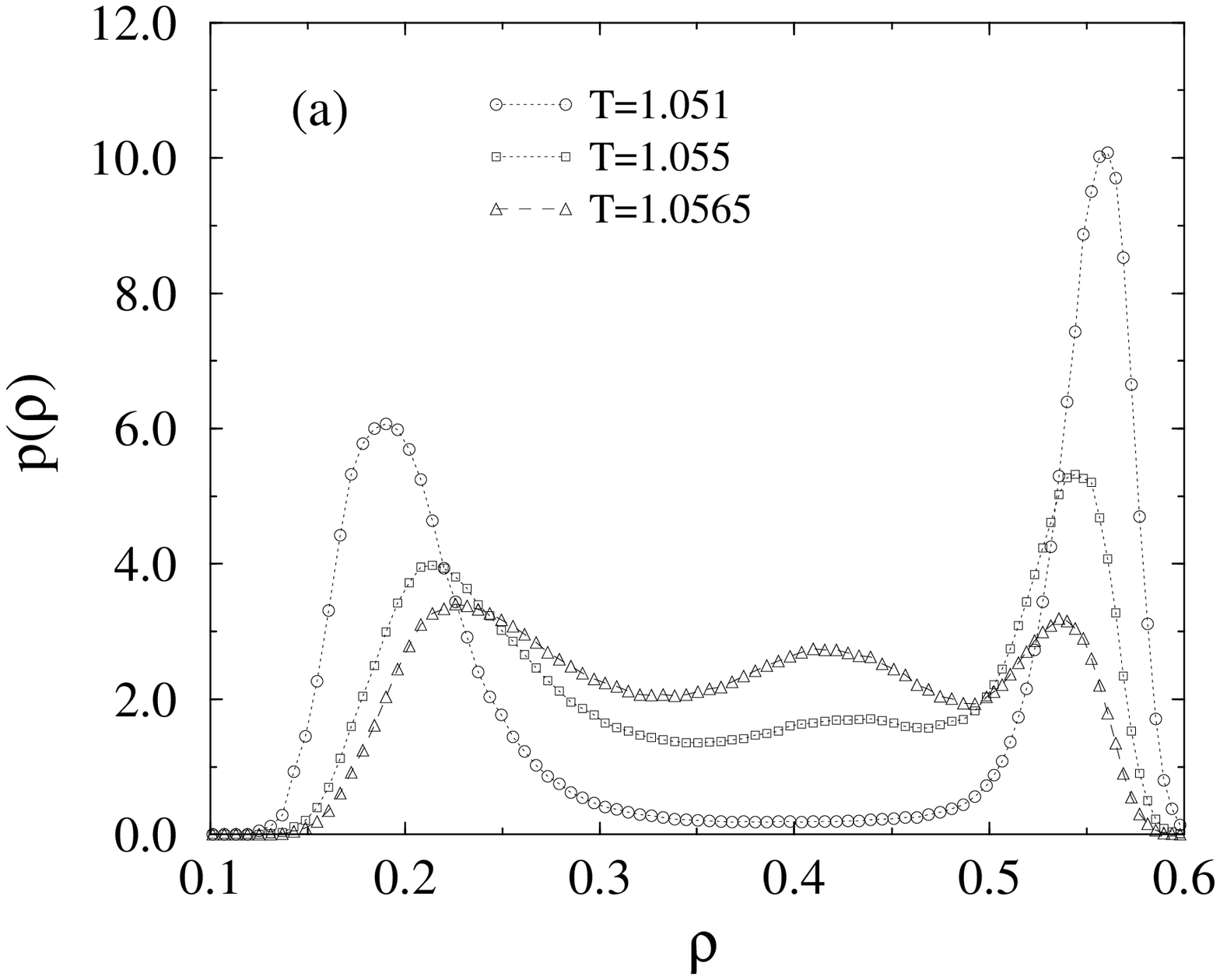}}} 
\centerline{\mbox{\epsffile{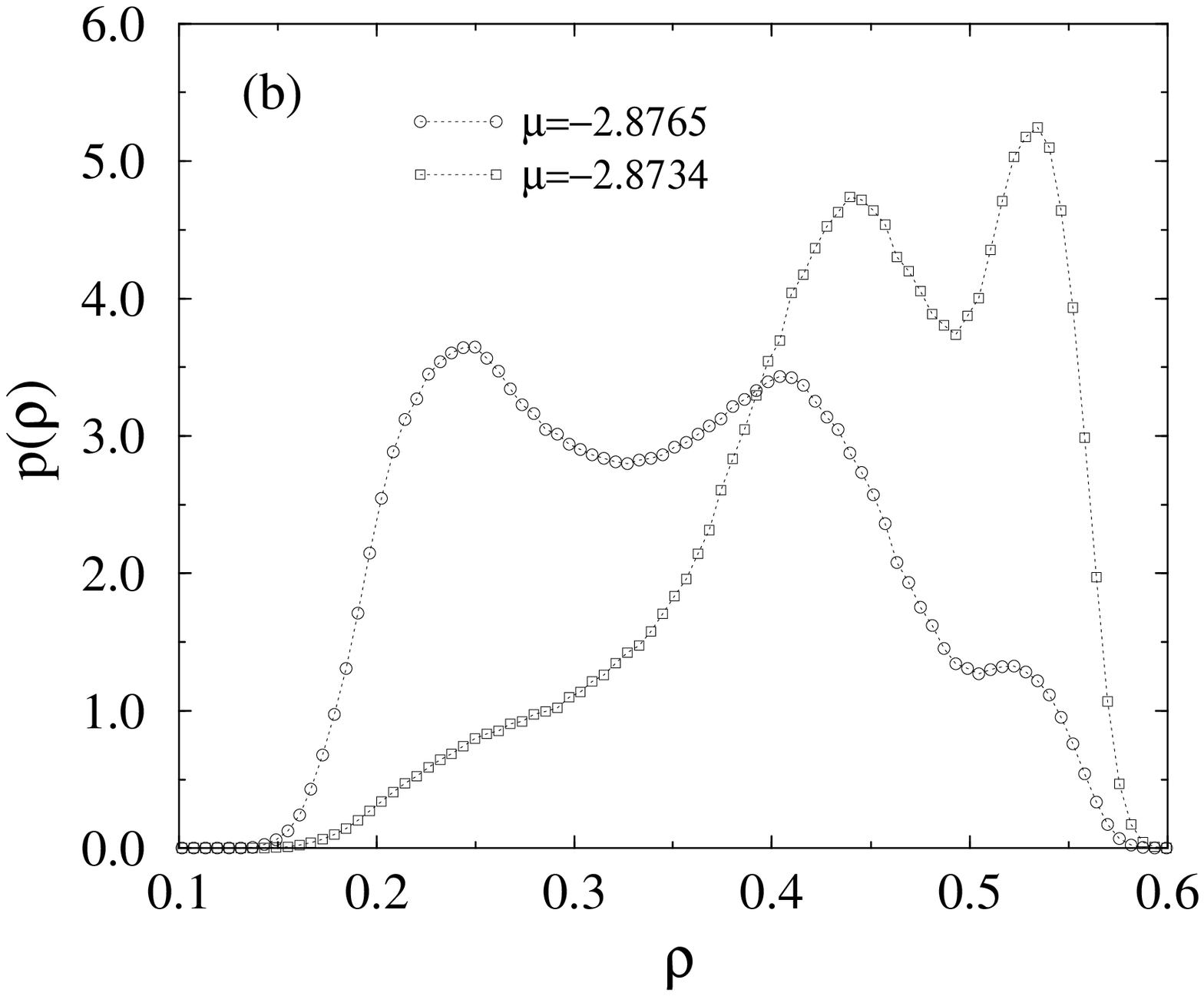}}} 
\centerline{\mbox{\epsffile{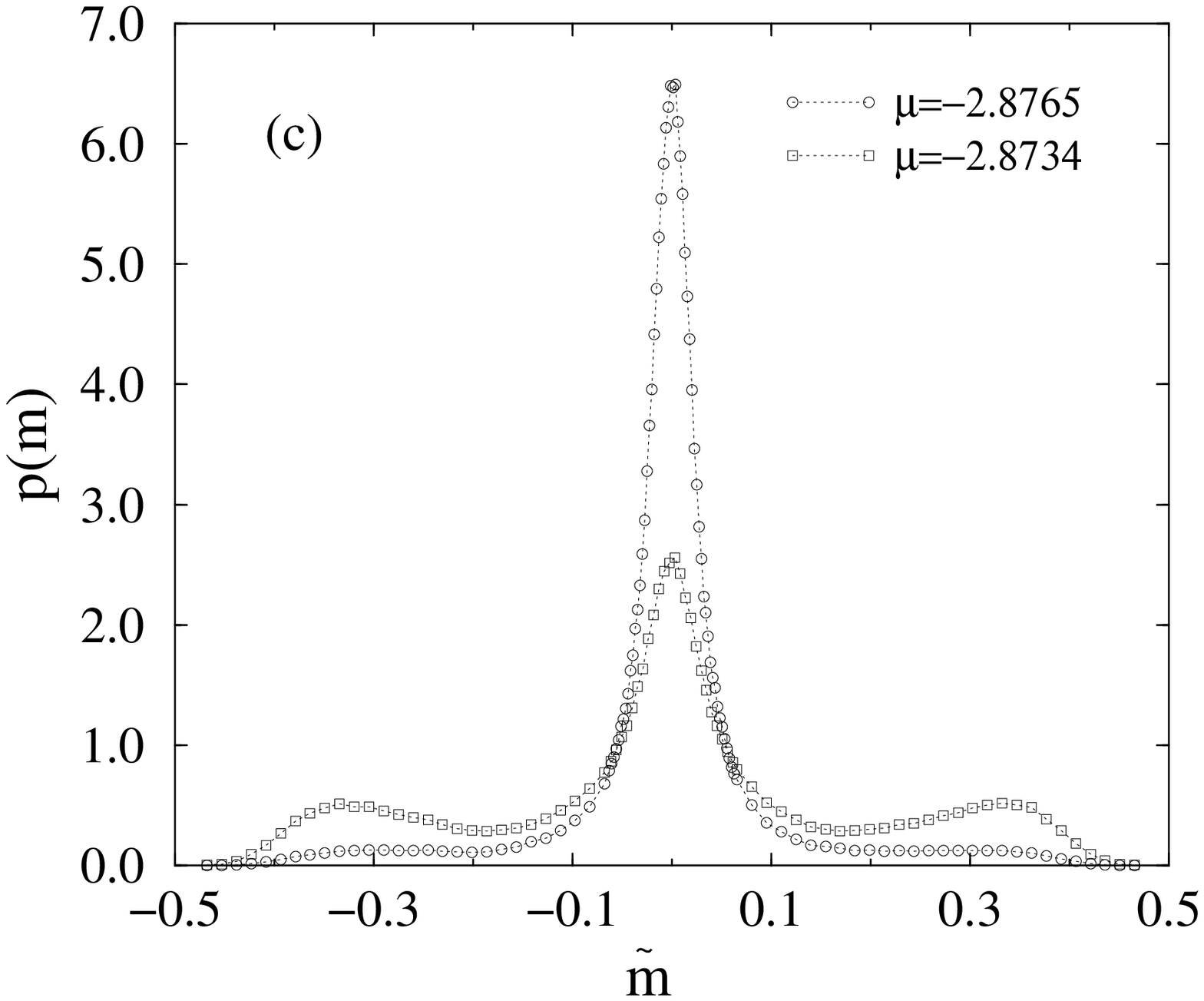}}} 

\caption{{\bf(a)} Selected coexistence density distributions for
$\delta=0.66$, spanning the triple point. {\bf (b)} The density
distribution for $T=1.058$ for two different values of the chemical
potential. {\bf (c)} The corresponding forms of $p(\tilde{m})$, where
$\tilde{m}=m\rho=(N_A-N_B)/V$. Lines are merely guides to the eye and
statistical errors are comparable with the symbol sizes.}

\label{fig:branch}
\end{figure}

The coexistence density distributions for $\delta=0.66$ are shown in
fig~\ref{fig:branch}a for temperatures spanning the triple point
temperature. For this $\delta$, the tricritical point is sufficiently
well separated from the LV line that it is possible to distinguish the
liquid-vapour and liquid-liquid branches by appropriately tuning the
chemical potential. This is demonstrated in fig.~\ref{fig:branch}b
which shows $p(\rho)$ for the two coexistence curves at $T=1.058$. The
different degree of order in the two liquid phases is clearly seen in
the distribution $p(\tilde{m})$, shown in fig.~\ref{fig:branch}c. One
notices, however, that both the density distributions show signs of the
third phase. This reflects the closeness of the two coexistence curves
at this $\delta$ and $T$, as evidenced by the very small chemical
potential difference. Under such conditions, finite-size smearing
effects render it difficult to completely isolate two of the three
phases.

\begin{figure}[h]
\vspace*{0.1 in}
\setlength{\epsfxsize}{7cm}
\centerline{\mbox{\epsffile{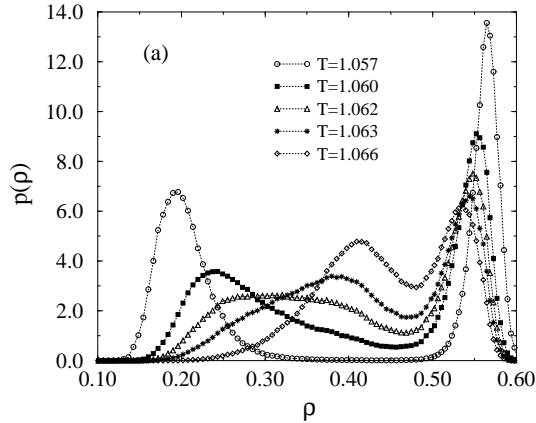}}} 
%\centerline{\mbox{\epsffile{pm_d0.65.eps}}} 

\caption{{\bf } Selected near-coexistence density distributions for
$\delta=0.65$ at a number of sub-tricritical temperatures.  Statistical
errors are comparable with the symbol sizes. }

\label{fig:tricrit}
\end{figure}

Finally in this section, we consider the phase behaviour for
$\delta=0.65$. Coexistence forms of $p(\rho)$ at selection of
temperatures are shown fig~\ref{fig:tricrit}. One observes that on
increasing temperature, the low density vapour peak moves smoothly over
to merge with the high density peak of the ordered liquid. At no point
is a three-peaked structure visible. This scenario is consistent with
the phase behaviour shown schematically in fig.~\ref{fig:bmf}d, in
which the vapour and the demixed liquid phases merge at a tricritical
point.

\section{Discussion and conclusions}
\label{sec:concs}

%Is there any experiment with which to compare?

In summary, we have used multicanonical Monte Carlo simulations and
histogram reweighting techniques to study how the liquid-vapour phase
behaviour of a symmetrical binary mixture depends on $\delta$, the
ratio of interaction strengths for dissimilar and similar particle
species. For $\delta\lesssim 1$, the phase diagram exhibits a critical
end point at temperatures well below the liquid-vapour critical point.
Decreasing $\delta$ shifts the critical end point closer to the
liquid-vapour critical point. For $\delta\approx 0.68$, however, the
critical end point becomes locally unstable, and a triple point occurs
in which vapour, a mixed liquid, and a demixed liquid all coexist. For
temperatures above the triple point there is coexistence either between
a high density demixed fluid and a moderate density mixed fluid, or
between a mixed fluid and its vapour. Decreasing $\delta$ still further
pushes the triple point to higher temperature, until for $\delta<0.65$
it eventually equals that of the isotropic liquid-vapour critical
point. Thereafter, the mixed liquid phase is preempted by the demixed
liquid phase and the liquid-vapour coexistence curve terminates in a
tricritical point.

Thus our simulation results confirm the qualitative picture of phase
diagram topology emerging from mean field theory as set out in
sections~\ref{sec:back} and \ref{sec:mf}.  These theories seem quite
successful in capturing key features of the behaviour such as the
existence of a coexistence curve anomaly at the CEP, the existence of
the triple point regime, and the crossover to a purely tricritical
regime. Additionally, our Landau theory study of
subsection~\ref{sec:landau} provides useful physical insight into the
manner in which the coupling of density and concentrations leads to the
observed phase behaviour.

In quantitative terms, however, the mean field theories are less
reliable. Owing to the neglect of correlations, they predict neither
the correct exponents for the coexistence curve singularities at the
CEP, nor the shape of the near critical LV coexistence curve.  The
values they yield for quantities such as the LV critical temperature
are also at variance with simulation estimates: e.g. for $\delta=0.72$
mean field calculations predict that the LV critical temperature is
$T^{mf}_{c0}=1.172$, while simulation gives $T^{sim}_{c0}=1.06(1)$. In
view of this, the apparently better agreement between the mean field and
simulation estimates of the CEP for $\delta=0.72$, {\em i.e.}
$T^{mf}_{CEP}=1.002$ and $T^{sim}_{CEP}=1.02(1)$, are presumable
fortuitous. For 3d tricritical behaviour, mean field theory is at least
expected  to yield the correct tricritical exponents, since the upper
critical dimension for such behaviour is $d=3$ \cite{LAWRIE}. Although
we have not attempted to probe the universal aspects of the tricritical
behaviour, our results show that estimates for the tricritical
temperature are not reproduced by the simulations, at least close to
the triple point regime. However, this may partly be the result of
crossover effects associated with the relative proximity of the LV
critical and tricritical points.

The mean field estimates are also inaccurate regarding the sensitivity
of the phase diagram topology to changes in $\delta$. The calculations
of section~\ref{sec:mfexp} predict that the regime of triple point
topology lies in the range $0.605<\delta<0.708$. In contrast, the
simulation results show this range to be considerably smaller, namely
$0.65\lesssim\delta\lesssim 0.68$. Indeed were it not for our use of
histogram reweighting to scan the phase behaviour as a function of
$\delta$, we might easily have missed this regime altogether. Thus it
seems that more sophisticated liquid state theories are called for
before the goal of accurately predicting the phase behaviour of simple
binary fluid models is attained. Presumably any successful theory must
be capable of dealing both with the critical and non-critical regimes
of the phase diagram. Infact one such theory, the Heirarchical
Reference Theory \cite{PAROLA} has recently been proposed. It would be
interesting to see whether or not it accurately reproduces the phase
behaviour of the present model.

% Given the close similarities between the mean field predictions for the phase 
% behaviour of the present model and those of the spin and dipolar 
% fluids, we believe our results have wider relevance than just to the 
% symmetrical binary fluid model.

With regard to further work on this model, one particularly interesting
project would be to investigate the predicted existence of the hidden
binodal and associated metastable critical point. The occurence of
metastable critical points was first discussed by Cahn \cite{CAHN}, and
later found in lattice gas models by Hall and Stell \cite{HALL}. More
recently it has suggested that they occur in colloidal fluids  close to
the freezing line \cite{EVANS}, in dipolar fluids \cite{GROH}, and in
models for water \cite{POOLE}.  Since the present model offers a
computationally tractable system, it might usefully be employed as a
test-bed for studying the generic features of the metastable critical
point. This could feasibly be achieved by quenching the system from high
temperature into the unstable regime just below the metastable critical
point. As described in section~\ref{sec:landau}, this should result in
a two-stage demixing process in which the metastable mixed liquid phase
appears for a transitory period before eventually demixing at later
times.

Additional interesting work would be to look at the equilibrium phase
behaviour of the symmetrical mixture as a function of $\delta<0$.
Landau theory \cite{ROUX} predicts that as $\delta$ is made
increasingly negative, the tricritical point transforms first into a
double critical end point, before a critical end point  emerges on the
{\em vapour} side of the LV coexistence envelope. It would certainly be
worthwhile to assess whether or not this scenario is correct.

\subsection*{Acknowledgements}

NBW thanks A.Z. Panagiotopolous and G. Stell for helpful discussions,
and M.E. Cates for bringing reference \cite{ROUX} to his attention. PN
thanks the DFG for financial support (Heisenberg foundation). This work
was supported by the EPSRC (grant number GR/L91412), the DFG (BMBF grant
Number 03N8008C), and the Royal Society of Edinburgh. Computer time on the
HLRZ at J\"{u}lich is also gratefully acknowledged.


\begin{thebibliography}{50}
\bibitem{NOTE1} The presence of a tricritical point in such a model is
traceable to its special symmetry.  In general, tricritical behaviour
cannot occur in a binary fluid mixture except by accident.
\end{thebibliography}
\end{document}